\documentclass[pre,superscriptaddress,unsortedaddress,showpacs,twocolumn,amsmath,amssymb]{revtex4}
\def\figwidth{0.75\linewidth}
\usepackage[final]{graphicx}
\def\A{{\text{A}}}
\def\B{{\text{B}}}

\let\bs\boldsymbol
\let\bf\relax
\let\bss\relax

\begin{document}
\title{Alpha-Relaxation Processes in Binary Hard-Sphere Mixtures}
\date{\today}
\def\tumphy{%
  \affiliation{Physik-Department, %
  Technische Universit\"at M\"unchen, 85747 Garching, Germany}}
\def\roma{%
  \affiliation{Dipartimento di Fisica and INFM Center for Statistical Mechanics
               and Complexity, Universit\`a di Roma ``La Sapienza'', Piazzale
               Aldo Moro 2, 00185 Roma, Italy}}
\author{G.~Foffi}\roma
\author{W.~G\"otze}\tumphy
\author{F.~Sciortino}\roma
\author{P.~Tartaglia}\roma
\author{Th.~Voigtmann}\tumphy

\begin{abstract}
Molecular-dynamics simulations are presented for two correlation
functions formed with the partial density fluctuations of binary
hard-sphere mixtures in order to explore the effects of mixing on the
evolution of glassy dynamics upon compressing the liquid into
high-density states. Partial-density-fluctuation correlation functions
for the two species are reported.  Results for the alpha-relaxation
process are quantified by parameters for the strength, the stretching,
and the time scale, where the latter varies over almost four orders of
magnitude upon compression.  The parameters exhibit an appreciable
dependence on the wave vector; and this dependence is different for
the correlation function referring to the smaller and that for the
larger species.  These features are shown to be in semi-quantitative
agreement with those calculated within the mode-coupling theory for
ideal liquid-glass transitions.
\end{abstract}

\maketitle

\section{Introduction}

If one compresses or cools a liquid, there appear slow dynamical processes
which are referred to as structural relaxation. These processes are
precursors of the liquid-glass transition. The study of these phenomena has
been a very active field of research in recent years. Several new
experimental techniques were introduced to measure the evolution of
structural relaxation spectra within the GHz band. Molecular-dynamics 
simulation techniques have been improved considerably so that correlation
functions of liquids in equilibrium can be obtained for time intervals covering
five to six orders of magnitude. The wealth of information obtained on glassy
dynamics is a challenge for the theory of amorphous condensed matter. However,
consensus on the understanding of the slow dynamics in glass-forming liquids
has not yet been achieved \cite{crete,pisa}.

Simple monoatomic liquids crystallize before structural relaxation dynamics is
fully developed. Therefore, studies of the glassy dynamics have to be performed
on simple molecular systems or suitable mixtures.
Recently, for example, a four-component mixture was studied by
neutron-scattering spectroscopy. This system transforms to a metallic glass
at low temperatures, but it exhibits the same scenario for the evolution of
structural relaxation as known for molecules \cite{Meyer2002}.
The first molecular-dynamics studies of structural relaxation in an
equilibrium liquid were performed for a binary mixture of particles
interacting by purely repulsive potentials \cite{Bernu1985,
Bernu1987,Roux1989}. A binary Lennard-Jones system
has been introduced \cite{Kob1994}, whose interaction potentials
are similar to the ones proposed for the description of the glass-forming
Ni-P mixture \cite{Stillinger1982}. This system has been used extensively to
analyze all facets of glassy dynamics in the equilibrium liquid and also
for the quenched non-equilibrium system \cite{Kob2003,Sciortino2001b}.
In the mentioned
previous studies, mixing was merely introduced as a means of suppressing
crystallization. In the present paper, we analyze the influence of mixing
on the structural relaxation.

In order to identify the effect of mixing on the glassy dynamics, we have
performed molecular-dynamics simulations for four binary hard-sphere mixtures
differing in the size ratio of the constituents and in the composition.
By increasing the total packing fraction up to $0.605$, the evolution of
structural relaxation was detected for a time interval up to five orders of
magnitude. As reported earlier \cite{Foffi2003}, two scenarios for mixing
effects have been identified. For a mixture with a small size disparity of
the constituents, the increase of the mixing percentage of the small particles
for a fixed total packing fraction leads to a slowing down of the long-time
dynamics. In this case, mixing stabilizes the glass state. However, upon
mixing particles with a large size disparity, the increase of the percentage
of the small particles at fixed packing fraction speeds up the structural
relaxation. In this case, mixing stabilizes the liquid. The present paper
reports a detailed analysis of the long-time relaxation processes,
traditionally referred to as alpha processes, for the two scenarios mentioned.

Glassy dynamics and a liquid-glass transition can also be observed
experimentally in colloidal
suspensions. In particular, one can prepare glass-forming
colloids where the interaction potential is a very good approximation to
a hard-sphere repulsion \cite{Pusey1991}. Informative light-scattering
studies of structural relaxation for such a hard-sphere suspension have
been reported \cite{Megen1993}. To suppress crystallization, a narrow
distribution of particle sizes was chosen. Strictly, such a system is a
multi-component mixture. But ignoring the small polydispersity, it can be
viewed as a one-component system. In the same sense, one can consider a colloid
studied by Henderson \textit{et~al.}\ \cite{Henderson1996} as an approximation
for a binary hard-sphere mixture with a size ratio $0.8$ between the two groups
of particles, and a mixture studied by Williams and van~Megen
\cite{Williams2001b} as one with size ratio $0.6$. For the first mixture,
a dramatic effect of mixing on the nucleation ratio was observed, but no
effect on the glassy dynamics has been reported. For the second mixture, it
was shown that the time scale for the alpha relaxation decreased upon mixing.
Our simulation results \cite{Foffi2003} suggest that the cited experiments
do not deal with colloid-specific features. Rather, they exemplify the two
scenarios for mixing effects on structural relaxation. Therefore, the
present paper provides a detailed list of quantitative predictions for
correlation functions of glassy colloids, which can be measured by
photon-correlation spectroscopy.

The mode-coupling theory for ideal liquid-glass transitions provides a
physical explanation for the evolution of structural relaxation in simple
systems and allows for the first-principle evaluation of the density-correlation
functions \cite{Goetze1992}. The results of this theory for the
hard-sphere system have been used for a detailed analysis of the
light-scattering data obtained for hard-sphere colloids with a small
polydispersity \cite{Megen1995}. The theory has been extended recently
to a discussion of binary hard-sphere mixtures. In particular, the
above mentioned two mixing scenarios had been obtained \cite{Goetze2003}.
It was possible to describe a major part of the scattering data for
a mixture \cite{Williams2001b} quantitatively by the theoretical results
\cite{Voigtmann2003pre}. These findings provide a motivation to use our
simulation results also for a detailed quantitative test of the
mode-coupling theory for the alpha-relaxation process.

The paper is organized as follows. In Sec.~\ref{sec.basic}, the simulation
details are described, representative results for the two mixtures considered
are exhibited, and the mode-coupling-theory formulas for the alpha-relaxation
process are listed. Then, in Sec.~\ref{sec.param}, it will be explained how
the alpha-relaxation processes are parameterized. The results for the
parameters are compared with the corresponding ones obtained from the
mode-coupling-theory findings. Section~\ref{sec.amaster} presents a comparison
of the alpha-relaxation master functions for the density-fluctuation
correlation functions of the simulation data with the corresponding theoretical
results. In Sec.~\ref{sec.conc}, the findings are summarized. 

\section{Basic concepts and results}\label{sec.basic}

\subsection{Specification of the systems}

Let $N$ and $\varrho$ denote the total number of particles and the total
number density, respectively, of the binary hard-sphere mixture (HSM)
to be studied. Further numbers specifying the system are the particle
diameters $d_\alpha$, the particle masses $m_\alpha$, their thermal
velocities $v_\alpha$, the partial number densities $\varrho_\alpha$ and
number concentrations $x_\alpha=\varrho_\alpha/\varrho=N_\alpha/N$,
as well as the
partial packing fractions $\varphi_\alpha=(\pi/6)\varrho_\alpha d_\alpha^3$.
Here, $\alpha=\A$ and $\B$ labels the big and small particles, respectively.
In the present work,
the size ratio $\delta=d_\B/d_\A$, the total packing fraction $\varphi=
\varphi_\A+\varphi_\B$, and the relative packing fraction of the smaller
species, $x=\varphi_\B/\varphi$, will be used as convenient control parameters
to characterize the thermodynamic state.

Density fluctuations of species $\alpha$ for wave vector $\vec q$ are
constructed from the positions $\vec r_k^{(\alpha)}$, $k=1,\ldots,N_\alpha$,
of particles of type $\alpha$: $\varrho_\alpha(\vec q)=\sum_k\exp[i\vec q
\vec r_k^{(\alpha)}]/\sqrt{N}$. The partial structure factors
$S_{\alpha\beta}(q)=\langle\varrho_\alpha(\vec q)^*\varrho_\beta(\vec q)
\rangle$ provide the simplest statistical information on the equilibrium
distribution of particles, i.e., on the structure. Here,
$\langle\cdot\rangle$ denotes canonical averaging. The structure factors
depend on the wave vector only via $q=|\vec q|$. They can
be written as $S_{\alpha\beta}(q)=x_\alpha\delta_{\alpha\beta}+\varrho
x_\alpha h_{\alpha\beta}(q)x_\beta$, with $h_{\alpha\beta}(q)$ denoting the
Fourier transform of the pair-correlation function. The latter can be
expressed in terms of the direct correlation functions $c_{\alpha\beta}(q)$
via the Ornstein-Zernike equation: $h_{\alpha\beta}(q)=c_{\alpha\beta}(q)
+\varrho\sum_\gamma c_{\alpha\gamma}(q)x_\gamma h_{\gamma\beta}(q)$
\cite{Hansen1986}. The $S_{\alpha\beta}(q)$, $h_{\alpha\beta}(q)$, and
$c_{\alpha\beta}(q)$ are elements of real symmetric two-by-two matrices.
The discussions will be restricted to such stable and metastable states where
$S_{\alpha\beta}(q)$ and $c_{\alpha\beta}(q)$ are smooth functions of $q$ and
of the control parameters $\varphi$, $\delta$, and $x$.

The main quantities of interest in this paper are the density correlators
$\Phi_{\alpha\beta}(q,t)=\langle\varrho_\alpha(\vec q,t)^*\varrho_\beta(\vec q)
\rangle$. These correlation functions provide the simplest statistical
characterization of the structural dynamics. They are real even
functions of time $t$, and they form the elements of a symmetric
two-by-two matrix. In principle, the correlators can be measured as
intermediate coherent scattering functions by neutron-scattering experiments
for conventional liquids, or by photon-correlation spectroscopy for colloidal
suspensions. A short-time expansion yields $\Phi_{\alpha\beta}(q,t)
=S_{\alpha\beta}(q)-(1/2)(qv_\alpha t)^2x_\alpha\delta_{\alpha\beta}
+{\mathcal O}(t^3)$ \cite{Hansen1986}. Within the regime
of normal-liquid states, say, $\varphi<0.4$, the short-time dynamics varies
on a $30\%$ level upon changes of the control parameters.
There are no structural relaxation phenomena apparent
in the transient dynamics.

Two mixtures shall be considered in the following. A system with
$\delta=0.60$ and $x=0.20$, referred to as the $\delta=0.60$-system,
is representative for a mixture with large size disparity. In this case,
$54\%$ of the particles are of species $\B$. A system with $\delta=0.83$
and $x=0.37$, referred to as the $\delta=0.83$-system, contains $50\%$
of small particles. It is representative for a mixture with small size
disparity. These mixtures have been used before, together with systems of
smaller percentages $x$, in order to demonstrate the evolution of mixing
anomalies with variations of $x$ \cite{Foffi2003}.
Since we are not interested in details of the short-time dynamics, the masses
of the particles are chosen equal, i.e.\ $v_\A=v_\B$.
The units of length and time are chosen such that $d_\A=1$ and
$v_\A=v_\B=1$. With these units, the natural time scale for the
microscopic motion is $t_{\text{mic}}=d_\A/v_\A=1$.

\subsection{Results from molecular dynamic simulations}

We perform standard molecular dynamics simulations for
binary mixtures of $N=1237$ and $N=700$ hard-sphere particles with size
ratios $\delta=0.60$ and $\delta=0.83$, respectively. The algorithm follows
the usual event-driven scheme for the simulation of hard-sphere particles
\cite{Rapaport1997}, where the trajectory of the system is propagated
from one collision to the next one. To generate dense enough initial
configurations without particle overlaps, we applied the same procedure as
described earlier \cite{Zaccarelli2002b}: Starting from a random distribution
of points, particles were separated growing their diameters in successive
steps until the desired size was reached. From the initial configuration,
each simulation proceeds by an equilibration run, followed by a production
run during which positions and velocities are saved for subsequent analysis.
In all cases, the equilibration time was larger than the time it takes for
the particles' average displacement to reach one diameter of the large
species, $d_\A$. Up to four independent runs per state point have been
performed to reduce statistical errors. Density correlation functions
$\Phi_{\alpha\beta}(q,t)$ and static structure factors $S_{\alpha\beta}(q)$
have been calculated by averaging over the independent runs and over $300$
different wave vectors $\vec q$ of the same modulus $q$.
The longest simulation run requested about three weeks, i.e.\ the largest
density studied took about three months of CPU time on a fast AMD Athlon
processor to be completed.

We checked that no crystallization occurred during the production runs
by monitoring the time evolution of the pressure of the system, and by
visual inspection of the configurations. We also evaluated the wave-vector
resolved structure factor without angular averaging to make sure that no
crystalline peaks have developed. Other mixture compositions than the ones
presented below have been tried; for $\delta=0.83$ and $x=0.276$ as well as
for $\delta=0.60$ and $x=0.10$, it was also possible to study the glassy
dynamics in the liquid phase \cite{Foffi2003}, despite a stronger tendency to
crystallization. A system with $\delta=0.60$ and $x=0.05$ did not stay in
the homogeneous liquid phase long enough for a study of structural relaxation.

\begin{figure}
\includegraphics[width=\figwidth]{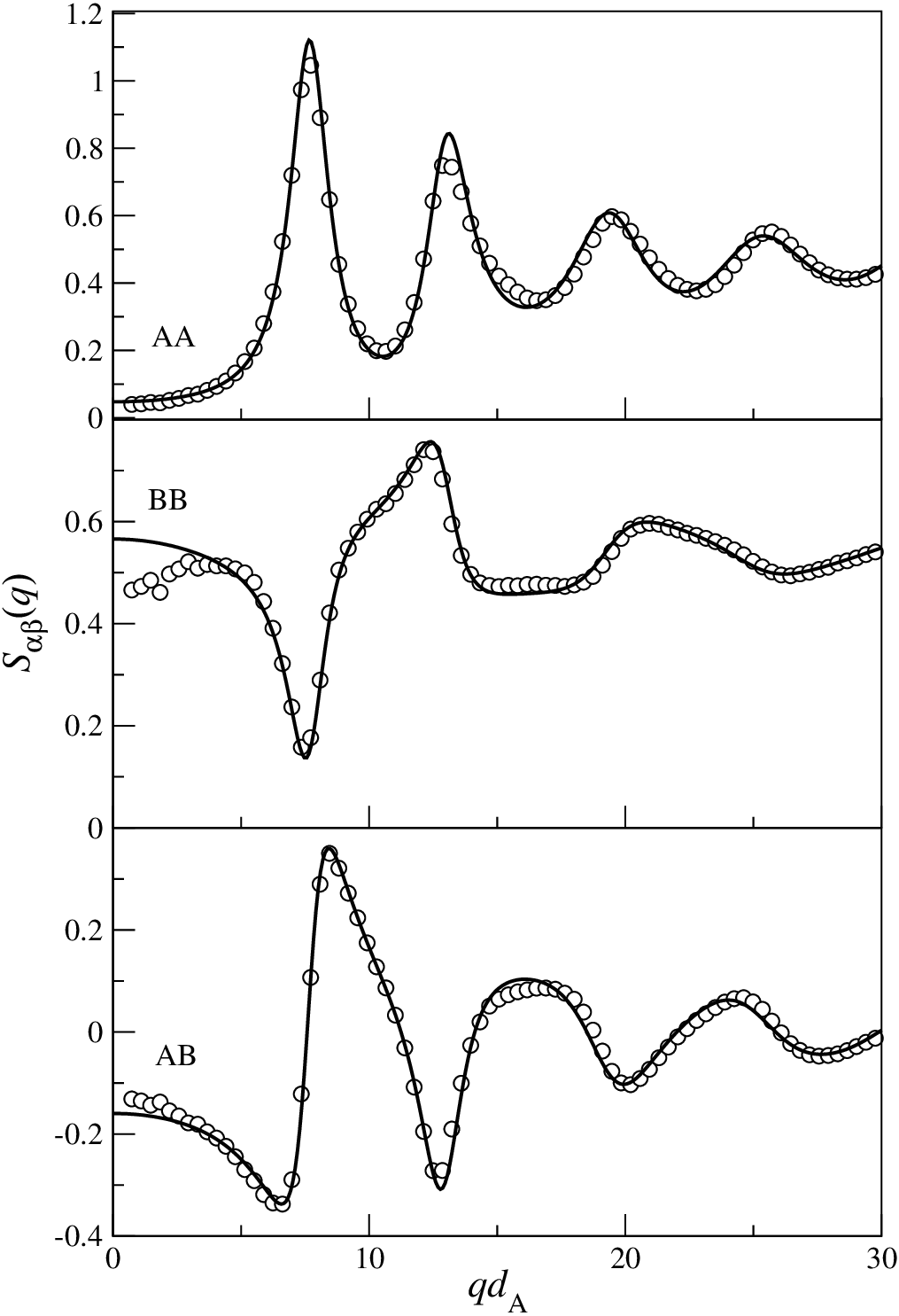}
\caption{\label{figsq}
  Partial structure
  factors $S_{\alpha\beta}(q)$, $\alpha,\beta=\A,\B$, of a binary hard
  sphere mixture (HSM) for diameter ratio of the particles
  $\delta=d_\B/d_\A=0.60$, a relative packing ratio $x$ of the smaller species
  of $20\%$, and total packing fraction $\varphi=0.60$. Circles are the
  molecular-dynamics simulation results, while full lines are
  the results calculated within Percus-Yevick theory.
  Here and in the following figures, the diameter of the larger
  spheres is chosen as the unit of length, $d_\A=1$.
}
\end{figure}

\begin{figure}
\includegraphics[width=\figwidth]{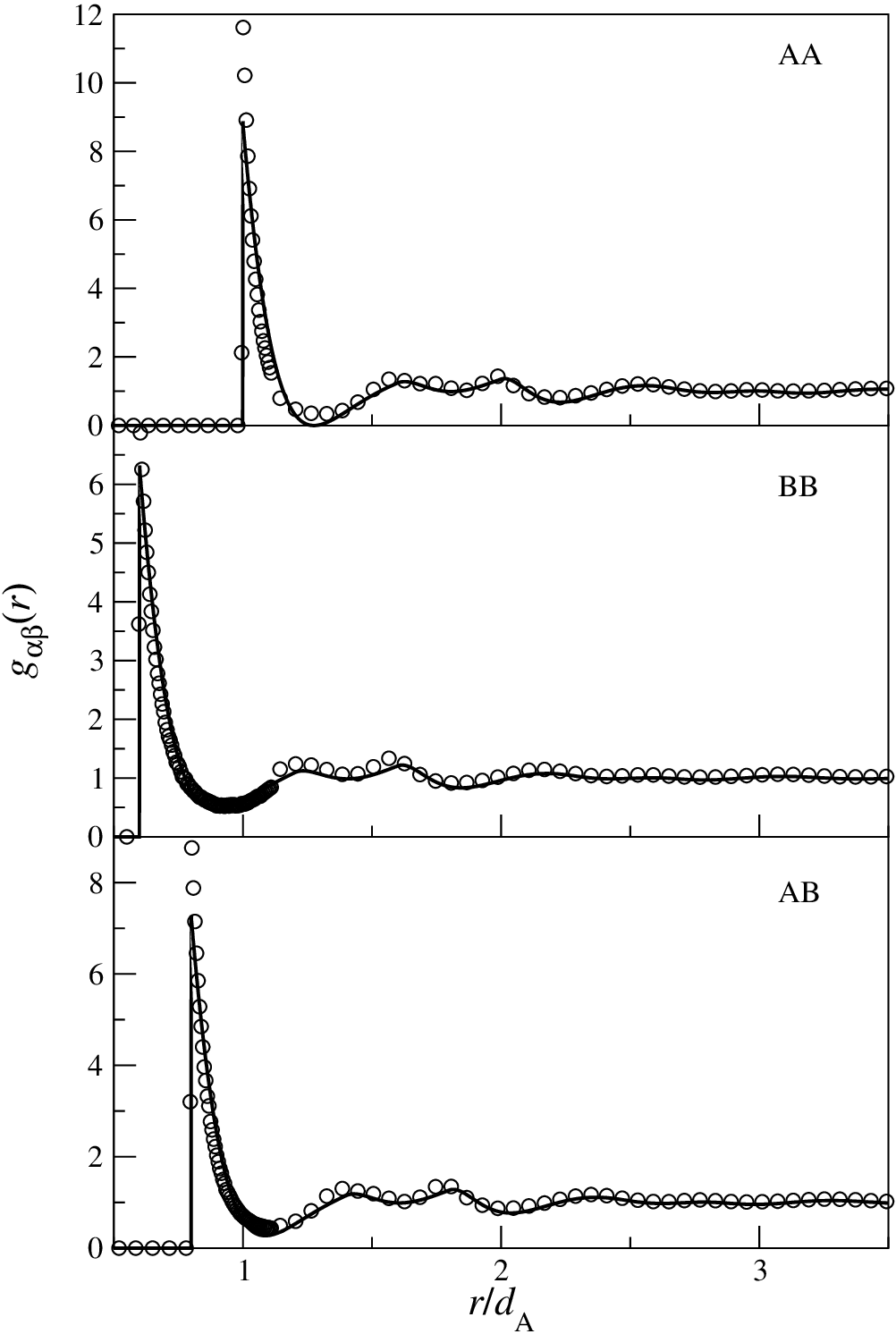}
\caption{\label{figgr}
  The pair distribution functions $g_{\alpha\beta}(r)$ corresponding to
  the results shown in Fig.~\protect\ref{figsq}.
}
\end{figure}

Figures \ref{figsq} and \ref{figgr} exhibit a typical set of structure
factors and the corresponding pair distribution functions. The results
refer to the $\delta=0.60$ system, and the lines are calculated using
the Percus-Yevick theory for HSM
\cite{Lebowitz1964b,Baxter1970}. Obviously, this approximation theory
accounts for the data rather well, even though the packing fraction
$\varphi=0.60$ considered is rather large.  But there are small
systematic deviations of the kind known from the discussion of
hard-sphere mixtures at smaller packing fractions
\cite{Malijevsky1997}. For example, the theory overestimates the
height of the first and second peak of $S_{\A\A}(q)$ by about
$10\%$. The contact values for the radial distribution function are
$11.6$, $6.87$, and $8.76$ for the $\A\A$, $\B\B$ and $\A\B$ function,
respectively, while the Percus-Yevick theory yields $8.88$, $6.32$,
and $7.28$, respectively. It will be discussed below that these
discrepancies have to be acknowledged if one intends to consider the
results of the mode-coupling theory (MCT) quantitatively. Even though
the results of Percus-Yevick theory are well-known, a side remark on
its qualitative features might be in order. Increasing the size
disparity, i.e.\ decreasing $\delta$ below unity, the height of the
first diffraction peak in $S_{\A\A}(q)$ decreases. Simultaneously, the
wing of the peak at $qd_\A\approx9$ increases.  Within MCT, the first
trend stabilizes the liquid state, while the second trend stabilizes
the glass. The first trend dominates at the glass transition for small
$\delta$, the second one that for larger $\delta$ \cite{Goetze2003}.
For the radial distribution functions, these trends correspond to a
reduction of the contact values and a decrease of the averaged radius
of the first neighbor shell with decreasing $\delta$. Moreover, at
larger size disparities, a splitting of the first-neighbor-shell peak
into a double peak is observed, as is evident in Fig.~\ref{figgr}.

\begin{figure}
\includegraphics[width=\figwidth]{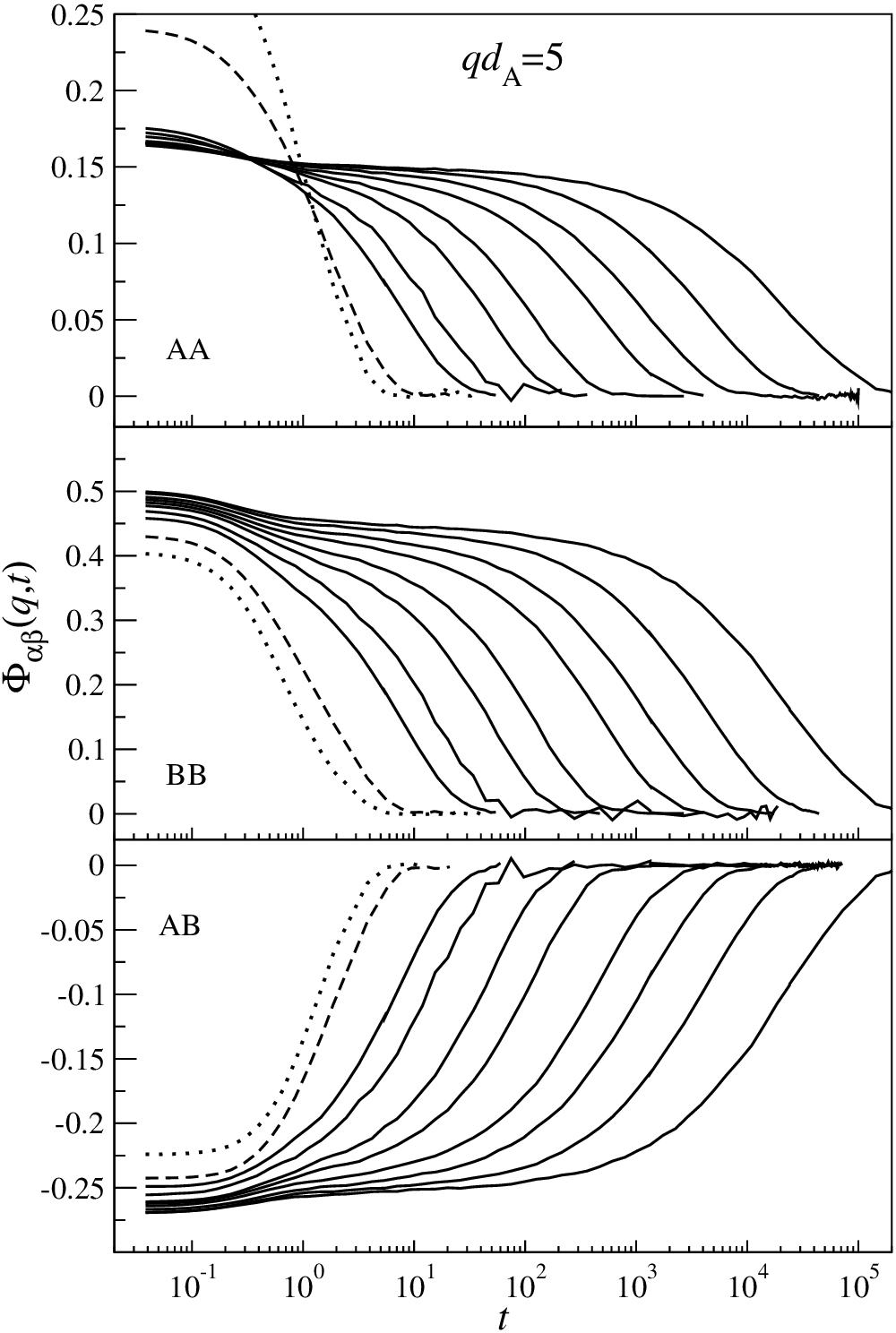}
\caption{\label{figphia}
  Molecular-dynamics-simulation results for the density correlators
  $\Phi_{\alpha\beta}(q,t)$, $\alpha,\beta=\A,\B$, for the $\delta=0.60$
  mixture for wave vector $qd_\A=5$. The dotted and dashed lines refer to
  packing ratios $\varphi=0.40$ and $0.45$, respectively. The full lines are
  correlators for $\varphi=0.530$, $0.550$, $0.570$, $0.580$, $0.590$, $0.595$,
  $0.600$, and $0.605$ (from left to right).
  Here and in the following figures, the unit of time is chosen such that
  the thermal velocity of the particles is unity.
}
\end{figure}

\begin{figure}
\includegraphics[width=\figwidth]{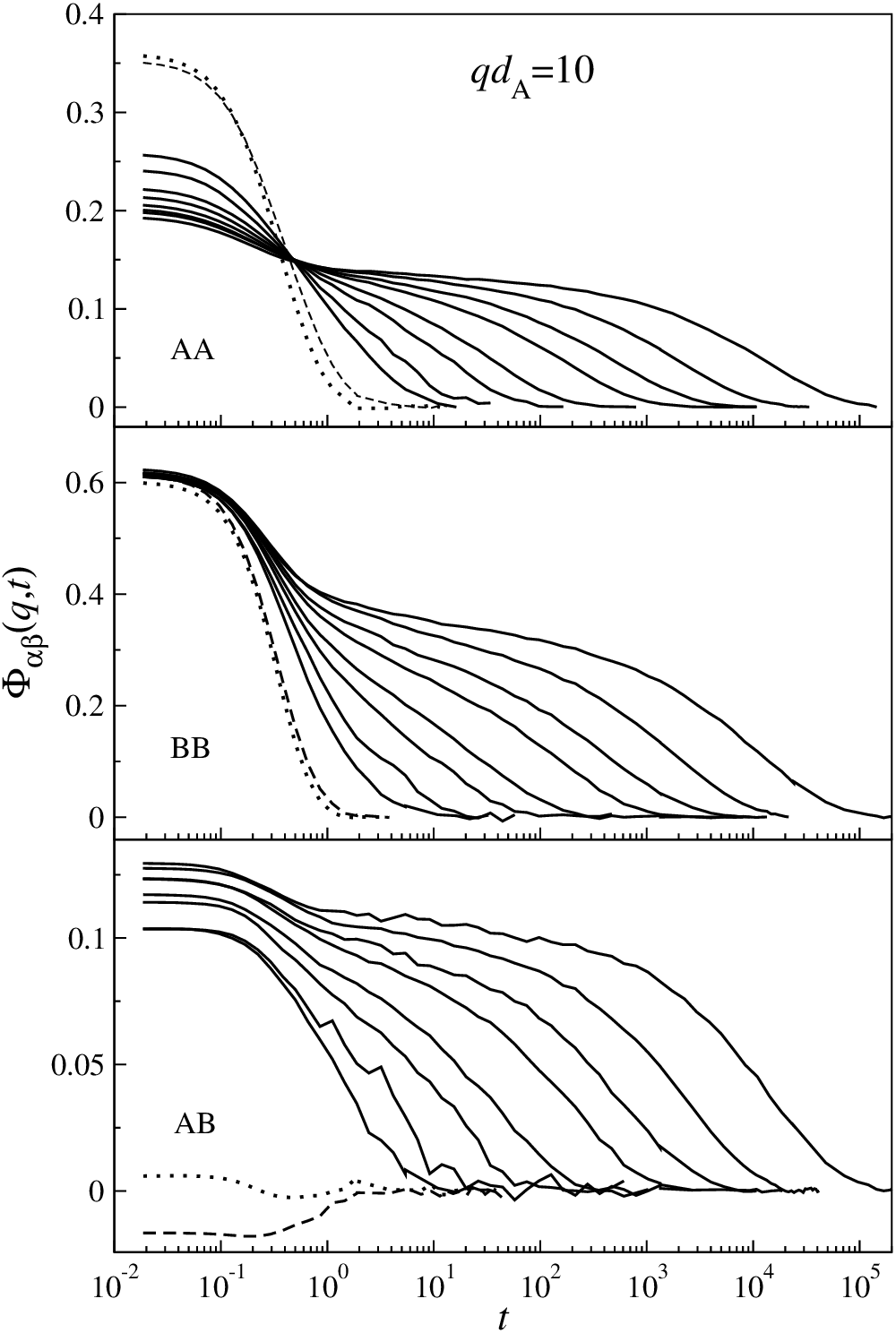}
\caption{\label{figphib}
  The analogous results as shown in Fig.~\protect\ref{figphia},
  but for wave number $qd_\A=10$.
}
\end{figure}

Figures \ref{figphia} and \ref{figphib} exhibit representative
examples for the evolution of the glassy dynamics upon compressing the
$\delta=0.60$ mixture. The wave vectors $qd_\A=5$ and $qd_\A=10$ have been
chosen since they exhibit the characteristic differences in mixing
effects for small and large wave vectors that have been discussed
before \cite{Foffi2003,Goetze2003}.  Note that the values of the
partial correlation functions can be quite different; in particular
the small values of the $\A\B$ correlator at $qd_\A=10$ are responsible
for the worse signal to noise ratio observed there.

The correlators for $\varphi=0.40$ are close to exponentials whose
characteristic decay time is near the natural time scale
$t_{\text{mic}}=1$ for the microscopic dynamics. They are typical for
normal-liquid behavior which can be described on a $30\%$ accuracy
level by Enskog's theory for dense gases
\cite{Boon1980}. If the packing fraction increases to $\varphi=0.45$, the time
scale for the decay of the correlators increases by about $40\%$, so that for
$t\approx5$ all correlators have decayed to below the $5\%$ level of their
initial values. Increasing $\varphi$ above $0.45$, a new relaxation pattern
evolves for the dynamics outside the transient regime, say for $t>5$. The
$\Phi_{\alpha\beta}(q,t)$-versus-$\log t$ diagrams exhibit a two-step
relaxation scenario that has repeatedly been observed before in simulation
studies and experiments. First, the correlators
decrease towards some plateau. The curves become flatter and the
plateau lengths increase if $\varphi$ increases. Then, the correlators
decrease from the plateau to zero. The dynamics for $t\gtrsim t_{\text{mic}}$
is called structural relaxation. Our simulations document this process, which
is characteristic for glass-forming liquids, for a time interval extending over
nearly five orders of magnitude.

The second step of the structural relaxation, i.e., the decay below the
mentioned plateau is conventionally referred to as the alpha process.
The figures demonstrate that the time scale for the alpha process increases
the faster with increasing $\varphi$ the larger the packing fraction. The
decay cannot be described by an exponential function, rather it is stretched
over wider time intervals. Obviously, the whole structural relaxation pattern,
in particular the alpha process, shows a subtle dependence on the wave vector
$q$. It is the goal of this paper to characterize the alpha process, in
particular its $q$ dependence, quantitatively.

\subsection{Some mode-coupling-theory equations}

Within MCT, the concept of a plateau and of an alpha-relaxation process can be
defined precisely in the sense of asymptotic laws describing the dynamics
near an idealized liquid-to-glass transition. These laws
provide a motivation for the parameterization of the data. In addition,
our data shall be used to test quantitatively the results of the theory. In
this section, the required formulas are compiled.

Let us introduce an obvious matrix notation to get the following equations in
a transparent form. $\bs S(q)$, $\bs\Phi(q,t)$, etc., shall denote
two-by-two matrices with elements $S_{\alpha\beta}(q)$, $\Phi_{\alpha\beta}
(q,t)$, etc. The Zwanzig-Mori formalism \cite{Hansen1986} can be used to
derive the exact equation of motion for the density correlators,
\begin{subequations}
\begin{equation}
  \bs I(q)\partial_t^2\bs\Phi(q,t) + \bs S^{-1}(q)\bs\Phi(q,t)
  +\int_0^t \bs M(q,t-t')\partial_{t'}\bs\Phi(q,t')\,dt'=\bs0\,.
\end{equation}
Here, $\bs I(q)$ is a matrix of inertia parameters,
$I_{\alpha\beta}(q)=1/(q^2v_\alpha^2x_\alpha)\delta_{\alpha\beta}$.
The kernels $M_{\alpha\beta}(q,t)$ are fluctuating-force correlation
functions, and they reflect the complicated many-body interaction effects.
The equation of motion has to be solved with the initial conditions
$\bs\Phi(q,t\!=\!0)=\bs S(q)$ and $\partial_t\bs\Phi(q,t\!=\!0)=\bs0$.
The essential step in the theory is the application of Kawasaki's
factorization approximation in oder to identify the kernel
contribution $\bs M^{\text{MCT}}(q,t)$, which expresses the coupling of the
forces to the density fluctuations. The remainder of the kernel,
$\bs M^{\text{reg}}(q,t)$, is assumed to describe the normal-liquid-state
dynamics. It is anticipated to vary regularly in the control parameters
and to decay on the scale $t_{\text{mic}}$ for the transient motion.
One gets $\bs M(q,t)=\bs M^{\text{reg}}(q,t)+\bs M^{\text{MCT}}(q,t)$,
where the mode-coupling kernel is given by the mode-coupling functional
$\mathcal F$,
\begin{equation} \bs M^{\text{MCT}}(q,t)
  ={\mathcal F}[\bs\Phi(t)](q)\,,
\end{equation}
which reads
\begin{widetext}
\begin{equation}\label{mctmemory}
  {\mathcal F}_{\alpha\beta}[\bs\Phi](q)=
  \frac1{2q^2}\frac{\varrho}{x_\alpha x_\beta}
  \sum_{\alpha'\beta'\alpha''\beta''}\sum_{\vec k}
  V_{\alpha\alpha'\alpha''}(\vec q,\vec k,\vec p)
  \Phi_{\alpha'\beta'}(k)\Phi_{\alpha''\beta''}(p)
  V_{\beta\beta'\beta''}(\vec q,\vec k,\vec p)\,.
\end{equation}
\end{widetext}
In the sum, the abbreviation $\vec p=\vec q-\vec k$ is used. The vertices
$V_{\alpha\alpha'\alpha''}(\vec q,\vec k,\vec p)$ are expressed in terms
of the direct correlation functions $c_{\alpha\beta}(k)$ and a triple
average $\langle\varrho_\alpha(\vec q)^*\varrho_\beta(\vec k)\varrho_\gamma
(\vec p)\rangle$. Simplifying the latter by the convolution approximation,
one gets
\begin{equation}\label{mctvertex}
  V_{\alpha\alpha'\alpha''}(\vec q,\vec k,\vec p)
  = (\vec q\vec k/q)c_{\alpha\alpha'}(k)\delta_{\alpha\alpha''}
  +(\vec q\vec p/q)c_{\alpha\alpha''}(p)\delta_{\alpha\alpha'}\,.
\end{equation}
Note that the mode-coupling functional is determined by the equilibrium
structure alone. Specifying the structure factors and a model for
$\bs M^{\text{reg}}(q,t)$, the preceding equations are closed. To
proceed towards a numerical solution, one introduces a grid of equally spaced
wave numbers extending up to a cutoff $q^*$. In this paper, we use
$140$ wave numbers and $qd_\A^*=56.8$ for the calculations based on the simulated
structure factors. Some results based on the Percus-Yevick
approximation shall also be shown, and
for those we used $200$ wave numbers up to $qd_\A^*=79.8$.
We refer to Ref.~\cite{Goetze2003} and the papers quoted there for further
details.
\end{subequations}

The MCT equations exhibit bifurcations for the long-time limits of the
solutions. For packing fractions $\varphi$ below some critical value
$\varphi^c=\varphi^c(\delta,x)$, one gets $\bs\Phi(q,t\to\infty)=\bs0$.
In this parameter regime, the solutions describe ergodic liquid dynamics.
For $\varphi\ge\varphi^c$, the long-time limits are non-degenerate symmetric
positive definite matrices $\bs\Phi(q,t\to\infty)=\bs F(q)$. For
these states, the solutions describe amorphous solids, i.e., ideal glasses.
The long-time limits obey the implicit equations
\begin{equation}\label{mctfeq}
  \bs F(q)=\bs S(q)-\left[\bs S(q)^{-1}+{\mathcal F}[\bs F](q)\right]
  ^{-1}\,.
\end{equation}
These equations are defined by the equilibrium structure alone; neither the
inertia matrix $\bs I(q)$ nor the regular memory kernel $\bs M^{\text{reg}}
(q,t)$ enter. The above equation for $\bs F(q)$ can be solved by a standard
iteration procedure \cite{Franosch2002}.

Let $\bs F^c(q)$ denote the non-degenerate positive definite matrix of
long-time limits at the transition point $\varphi=\varphi^c$. For reasons
of continuity, $\bs\Phi(q,t)$ has to be close to $\bs F^c(q)$ for a large
time interval if $|\varphi-\varphi^c|$ is small. The correlators are the
closer to $\bs F^c(q)$ the smaller $|\varphi-\varphi^c|$, and the time
interval of this close approach extends simultaneously. Thereby,
the evolution of the plateaus, which were discussed above in connection
with Figs.~\ref{figphia} and \ref{figphib}, is explained by MCT, and the
$\bs F^c(q)$ are the MCT expressions for the plateaus.

The decay of the correlators from the plateaus to zero for small negative
$\varphi-\varphi^c$ shall be characterized by some time scale $\tau(\varphi)$.
Obviously, $\lim_{\varphi\to\varphi^c}\tau(\varphi)=\infty$. Let us consider
the dynamics of the relaxation from the plateau on the time scale $\tau$
by writing $t=\tilde t\tau$ with $\tilde t$ fixed but positive. There holds
\begin{equation}\label{alphascaling}
  \lim_{\varphi\to\varphi^c}\bs\Phi(q,\tilde t\tau)
  =\bs\Psi(q,\tilde t)\,,
\end{equation}
where $\bs\Psi(q,\tilde t)$ obeys the equation \cite{Fuchs1993}
\begin{subequations}
\begin{equation}
  \bs\Psi(q,\tilde t)=\bs S(q)\tilde{\bs M}(q,\tilde t)\bs S(q)
  -\bs S(q)\frac{d}{d\tilde t}\int_0^{\tilde t}
   \tilde{\bs M}(q,\tilde t-\tilde t')\bs\Psi(q,\tilde t')\,d\tilde t'\,,
\end{equation}
to be solved with the initial condition $\bs\Psi(q,\tilde t\!=\!0)=\bs F^c(q)$.
Here, $\tilde{\bs M}(q,\tilde t)$ is determined by the mode-coupling
functional for the critical point,
\begin{equation}
  \tilde{\bs M}(q,\tilde t)={\mathcal F}^c[\bs\Psi(\tilde t)](q)\,.
\end{equation}
The numerical solution of the equation for $\bs\Psi(q,\tilde t)$ is done
similar to that for the full equations of motion.
\end{subequations}

Equation \eqref{alphascaling} implies the following conclusion. Given some
$\tilde t_-$ and some error margin,
\begin{equation} \bs\Phi(q,t)=\bs\Psi(q,t/\tau)\end{equation}
is valid within the margin for $\tilde t=t/\tau>\tilde t_-$, provided
$\varphi-\varphi^c$ is small enough. This is the superposition principle
for the MCT alpha process. It describes the correlators in terms of a
$\varphi$-independent master function $\bs\Psi(q,\tilde t)$, and attributes
the strong $\varphi$-dependence to that of the scale $\tau$. Presenting
the correlators as functions of $\tilde t$, the interval for $\tilde t$
where they coincide expands to lower values of $\tilde t$ if
$|\varphi-\varphi^c|$ decreases. The master functions $\bs\Psi(q,\tilde t)$
depend only on the equilibrium structure. Neither the inertia parameters
in $\bs I(q)$ nor the regular kernel $\bs M^{\text{reg}}(q,t)$ have any
influence on $\bs\Psi(q,\tilde t)$. These quantities enter the time scale
$\tau$ only.

There are complicated but straight-forward formulas to evaluate from
${\mathcal F}^c$ the so-called von~Schweidler exponent $b$, $0<b<1$, a
critical amplitude $\bs H^{(1)}(q)$, which is a positive definite matrix,
and a correction amplitude $\bs H^{(2)}(q)$ \cite{Voigtmann2003a}.
These quantities determine the von~Schweidler expansion of the master
functions,
\begin{equation}\label{vonschweidler}
  \bs\Psi(q,\tilde t)=\bs F^c(q)-\bs H^{(1)}(q)\tilde t^b
  +\bs H^{(2)}(q)\tilde t^{2b}\,.
\end{equation}
Here, terms of order $\tilde t^{3b}$ have been dropped.
Thereby an equation is obtained for the beginning of the alpha process.

The MCT equations have been studied before for binary HSM using the
Percus-Yevick approximation for the structure factors \cite{Goetze2003}.
In the present paper, results will be presented using the structure factors
$\bs S(q)$ obtained from the simulation work. For the two mixtures
we have calculated
\begin{subequations}\label{fit}
\begin{align}
  &\varphi^c_{\text{MCT}}=0.548,\,b=0.44&
    &\text{($\delta=0.60$, $x=0.20$);}\label{fit1}\\
  &\varphi^c_{\text{MCT}}=0.545,\,b=0.43&
    &\text{($\delta=0.83$, $x=0.37$).}\label{fit2}
\end{align}
\end{subequations}

From the simulation data, the critical packing fractions $\varphi^c$ for the
liquid-glass transitions of the two mixtures have been determined from the
alpha-relaxation times of the density auto-correlation functions, see
Sec.~\ref{secalphascale}, yielding values of $0.606$ and $0.586$, respectively.
The errors of $10\%$ and $7\%$, respectively, exhibited
by the values noted in Eqs.~\eqref{fit1} and \eqref{fit2}, indicate the
uncertainty one should expect for MCT results. It is worth to stress that
if one bases MCT on
Percus-Yevick approximation for the structure factors, one gets as critical
values for the two mixtures $0.520$ and $0.515$, respectively. Hence,
the use of correct instead of approximated structure factor input to
the theory improves the results for the critical points.
It is remarkable that the modest errors of the Percus-Yevick theory, which
are exhibited in Fig.~\ref{figsq}, lead to noteworthy changes in the MCT
results for the critical points.

\section{Parameterization of the alpha-relaxation processes}\label{sec.param}

\subsection{Evolution of the alpha process}

\begin{figure}
\includegraphics[width=\figwidth]{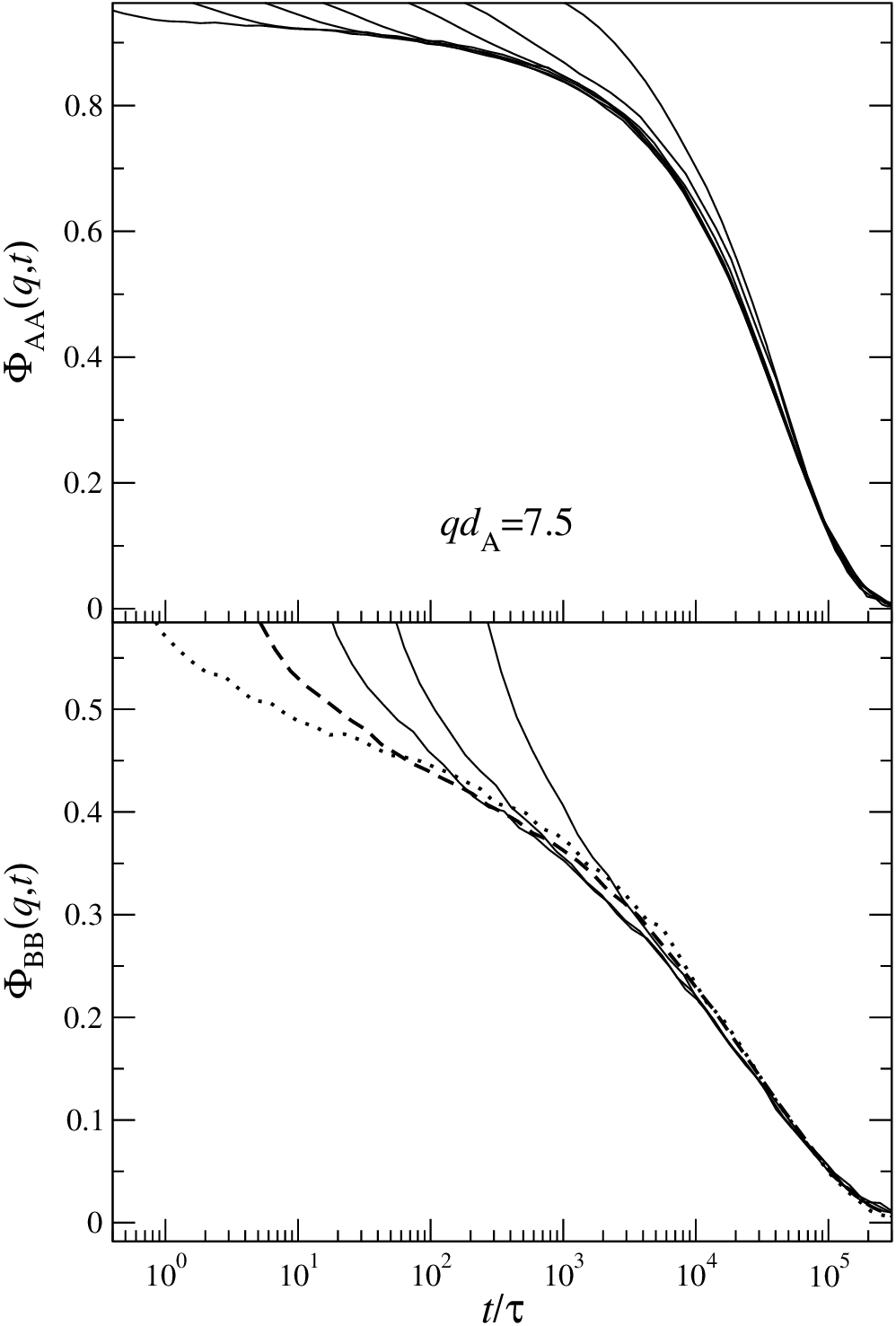}
\caption{\label{alphasc}
  Density auto-correlation functions $\Phi_{\A\A}(q,t)$ and
  $\Phi_{\B\B}(q,t)$ of the $\delta{\bf}{\bss=\bss d_\B/\bss
  d_\A}=0.60$ system for wave number $qd_\A=7.5$, presented as functions
  of the rescaled time $\tilde t=t/\tau$. The upper panel exhibits
  simulation results for packing fractions $\varphi=0.605$, $0.60$,
  $0.595$, $0.59$, $0.58$, $0.57$, and $0.53$; the lower panel for
  $\varphi=0.605$, $0.60$, $0.595$, $0.59$, and $0.58$ (from left to
  right).  For each $\varphi$, the scaling time $\tau$ is chosen such
  that the curves coincide for large times.  }
\end{figure}

The evolution of the alpha-relaxation scaling law is examined in
Fig.~\ref{alphasc}. The upper panel is a typical example for the
majority of correlators obtained in our simulations. For every packing
fraction $\varphi$, some time scale $\tau\equiv\tau(\varphi)$ can be defined
so that the long-time parts of the correlators coincide if these are
considered as functions of $\tilde t=t/\tau$. This coinciding part
provides the master functions $\Psi_{\alpha\beta}(q,\tilde t)$ for the
alpha process of the fluctuation considered. Upon increasing $\varphi$,
the $\tilde t$ interval where Eq.~\eqref{alphascaling} holds, expands to
lower values of the rescaled time $\tilde t$. Thus, the observed scenario
confirms the MCT prediction.
However, our data also exhibit violations of the above described scenario,
which cannot be understood in the framework of MCT. These occur only for the
$\B\B$ correlators
of the $\delta=0.60$ mixture for wave vectors around the structure-factor
peak position, $qd_\A\approx7$; and this only for the two largest packing
fractions examined, $\varphi=0.600$ and $0.605$. The lower panel of
Fig.~\ref{alphasc} shows a representative example. The following
analysis of $\bs\Psi(q,\tilde t)$ shall therefore be based on those data
sets that do not exhibit the described phenomenon; i.e.\ for $\varphi<0.60$.

\begin{figure}
\includegraphics[width=\figwidth]{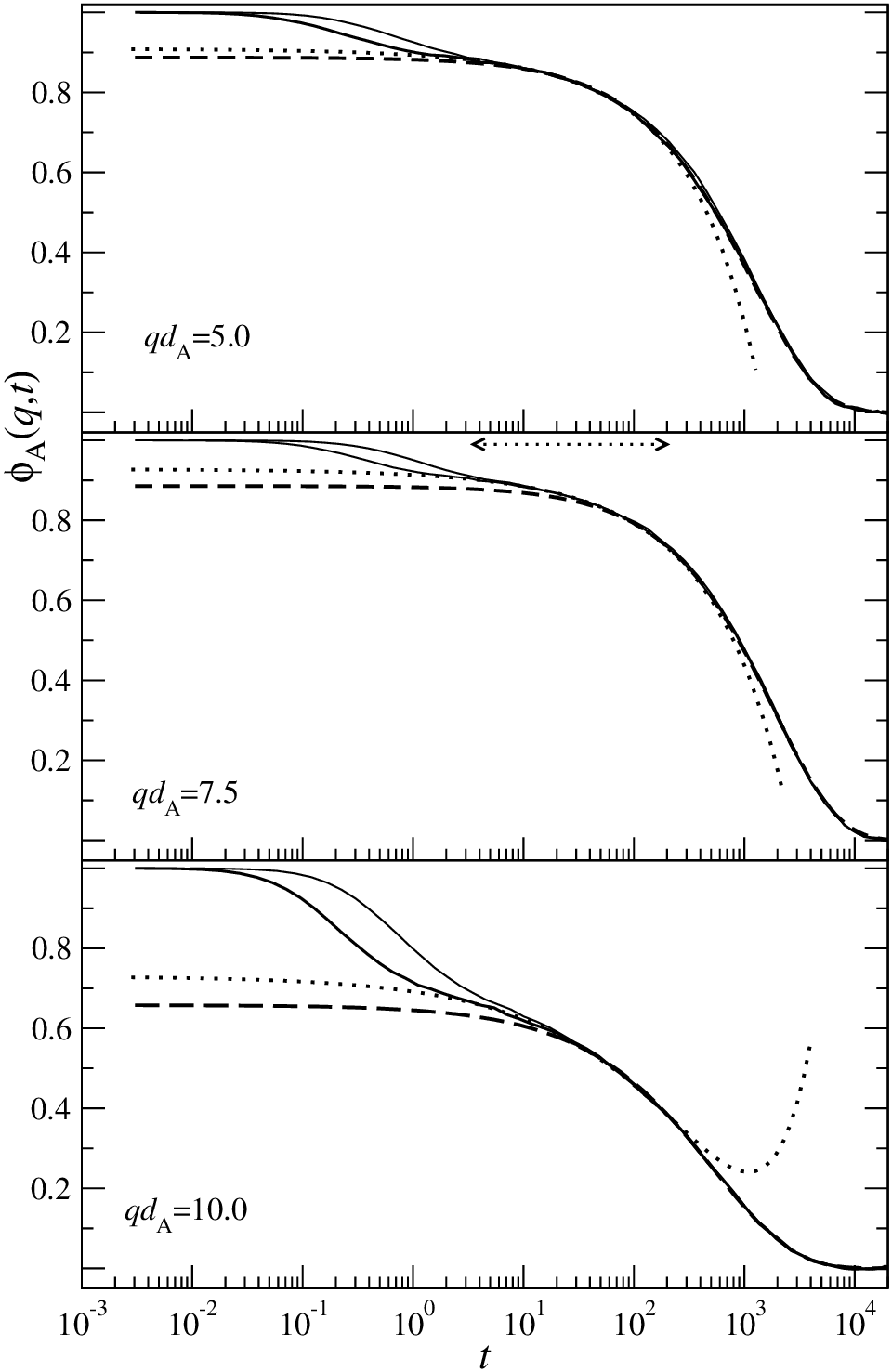}
\caption{\label{mdkww}
  Normalized large-particle density auto-correlation functions
  $\phi_\A(q,t)=\Phi_{\A\A}(q,t)/S_{\A\A}(q)$ obtained from the
  simulations for three wave numbers $q$ and packing fraction
  $\varphi=0.595$ (left full line) for the $\delta=0.60$ mixture.  The
  right full lines are the simulation results for $\varphi=0.590$,
  rescaled on the first curve for large times $t$. The dashed and
  dash-dotted lines are fits to the alpha process by the Kohlrausch
  function, Eq.~\protect\eqref{kww}, and the von~Schweidler expansion,
  Eq.~\protect\eqref{vonschweidler}, with $b=0.44$, respectively. The
  fit interval for the latter is indicated by the horizontal dotted
  arrow.}
\end{figure}

\begin{figure}
\includegraphics[width=\figwidth]{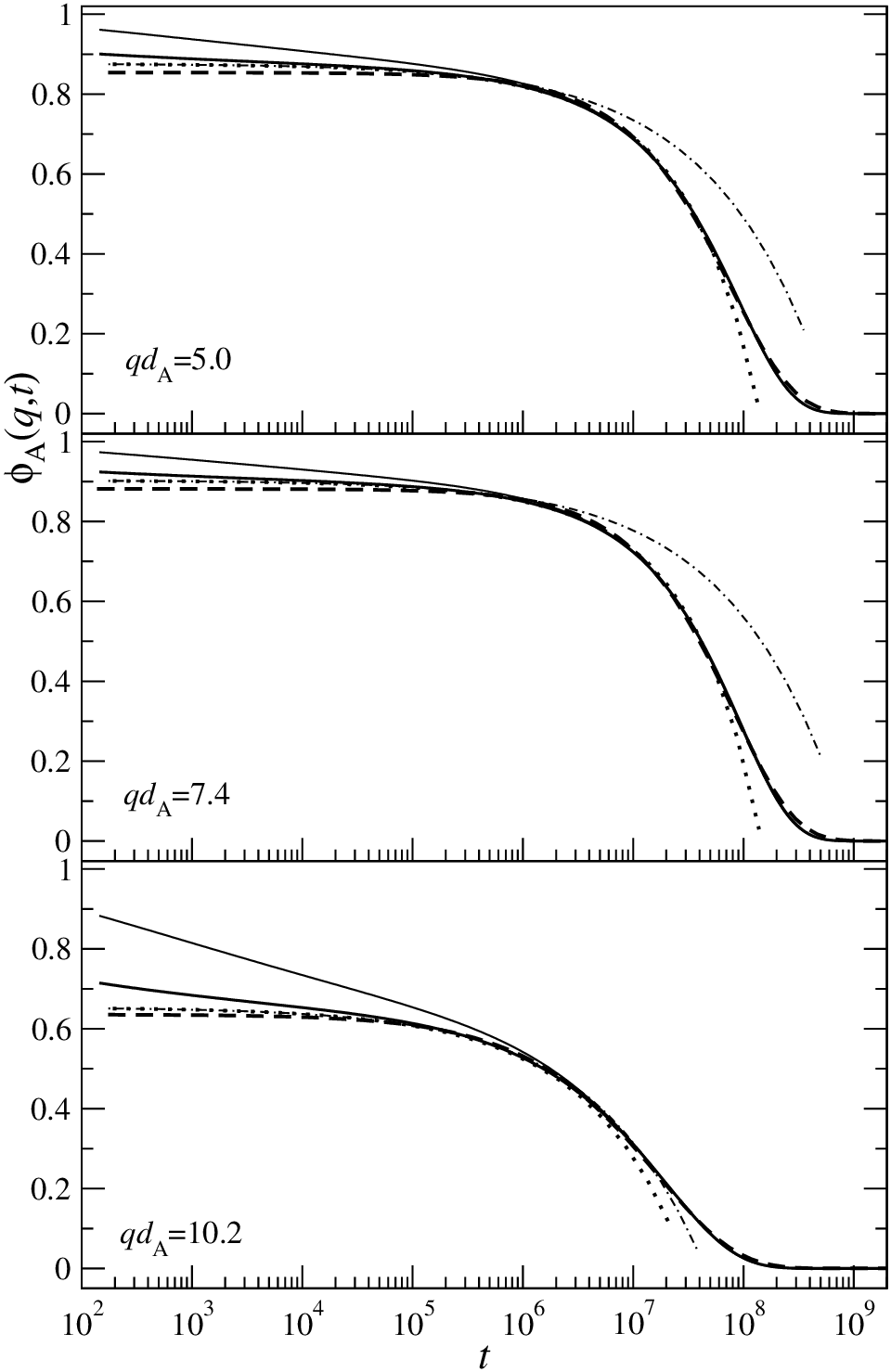}
\caption{\label{mctkww}
  The analogous set of curves as in Fig.~\protect\ref{mdkww}, but now
  obtained for the numerical solutions of the MCT equations. The packing
  fractions for the full curves are $\varphi=0.548$ and $0.545$. The
  additional dash-dotted lines are the asymptotes given by the first two terms
  in Eq.~\protect\eqref{vonschweidler}.
}
\end{figure}

The stretched exponential,
\begin{equation}\label{kww} \phi_\alpha^{\text{K}}(q,\tilde t)
  =A_\alpha(q)\exp\left[-(\tilde t/\tilde\tau_\alpha(q))^{\beta_\alpha(q)}
  \right]\,,
\end{equation}
is an often used empirical function for the description of
alpha processes. It was introduced by Kohlrausch for the description of
dielectric relaxation data. The description of the alpha-process master
function by an amplitude---also called plateau value---$A_\alpha(q)$, a time
scale $\tilde\tau_\alpha(q)$,
and a Kohlrausch exponent $\beta_\alpha(q)$ shall be used here as well.
The dashed lines
in Fig.~\ref{mdkww} exhibit representative examples for an analysis
of the normalized auto-correlation functions
$\phi_\A(q,t)=\Phi_{\A\A}(q,t)/S_{\A\A}(q)$. The
figures contain rescaled data for $\varphi=0.595$ and $0.590$, in oder to
identify a major part for the interval of rescaled times $\tilde t$,
for which the superposition principle is valid. Similarly, Fig.~\ref{mctkww}
exhibits examples for a fit of the stretched exponentials to the
numerical solutions of the MCT equations. The $q$ vectors are as close to
the ones of Fig.~\ref{mdkww} as permitted by the use of discrete
wave vector grids. Note that the choice of the overall time scale is
irrelevant for the discussion of the master functions $\bs\Psi(q,\tilde t)$.

Figures \ref{mdkww} and \ref{mctkww} confirm an observation often made
earlier: the fits by Eq.~\eqref{kww} provide a good description of a major
part of the alpha process. However, there are also systematic deviations
between the fit function $\phi_\alpha^{\text{K}}(q,\tilde t)$ and the master
functions $\Psi_{\alpha\alpha}(q,\tilde t)$.
This holds in a similar manner for the fits to the data and to the MCT
results. The fit contains unavoidable systematic errors, because the fit
parameters depend somewhat on the time interval chosen for the fit
optimization. In our analysis, the fit was done so that the large-$\tilde t$
part is described best. Thereby, the errors of the fit appear solely for the
small-$\tilde t$ part of the master functions.

Equation \eqref{vonschweidler} suggests another fit formula, which is valid for
the small-$\tilde t$ part of the master functions. But the small-$\tilde t$
part can be identified only to the limited extend to which the scaling
interval can be established. Consequently, also the fit using the
von~Schweidler series contains unavoidable uncertainties. The dotted lines
in Fig.~\ref{mdkww} exhibit representative examples for such fits.
Fits could be achieved with von~Schweidler exponents $b$ chosen between
$0.40$ and $0.50$. Therefore, the predicted exponents, Eq.~\eqref{fit},
are confirmed within an uncertainty of $\pm0.05$. All the results shown
are obtained with the cited theoretical exponents. The
remaining fit parameters shall be discussed below.

Note that these results can depend somewhat on the time window chosen
for the fit, which is $3.5<t<200$ for the fits discussed here.
Similarly, the dotted lines in Fig.~\ref{mctkww} exhibit the results
of Eq.~\eqref{vonschweidler} for the MCT results.  But here, the
functions $\bs F^c(q)$, $\bs H^{(1)}(q)$, and $\bs H^{(2)}(q)$, as
well as the scale $\tau$ are calculated from the MCT equations. In
this sense, the dotted lines are not fit results.  The discrepancies
between the dotted and the full lines represent the ones between the
full solution of the MCT equations of motion and the specified
second-order asymptotic description of the solution. It is reassuring
that the discrepancies between the full and the dotted lines in
Fig.~\ref{mdkww} exhibit similar trends as the ones shown in
Fig.~\ref{mctkww}. A quantitative account of the differences between
the results shown in the two figures is included in the discussions of
the following sections.

\subsection{The plateau values}

\begin{figure}
\includegraphics[width=\figwidth]{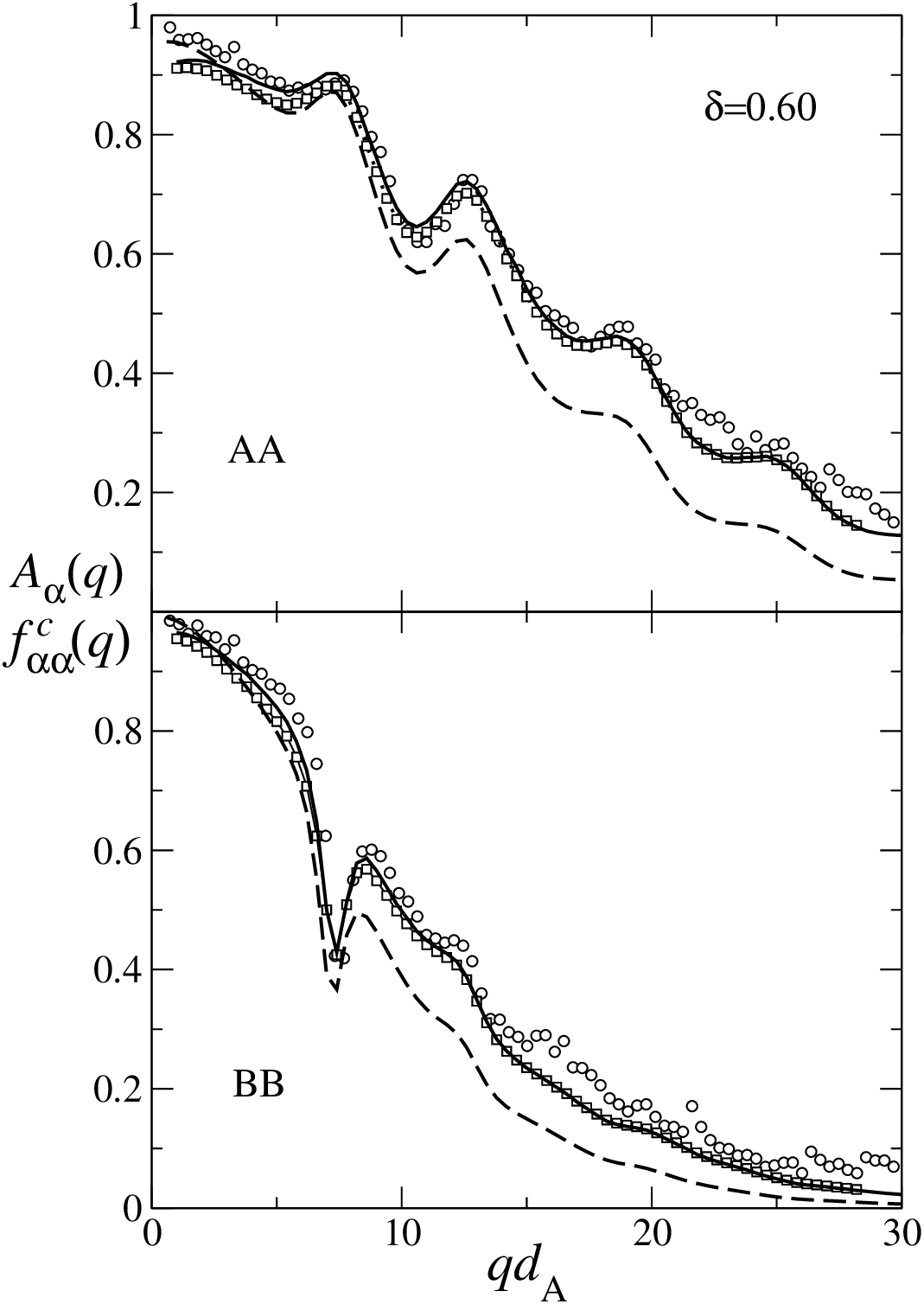}
\caption{\label{mdampl60}
  Circles and squares represent the Kohlrausch amplitudes $A(q)$ determined
  by fitting Eq.~\protect\eqref{kww} to the simulation data and MCT solutions,
  respectively, for the normalized auto-correlation functions
  $\Phi_{\alpha\alpha}(q,t)/S_{\alpha\alpha}(q)$ of the big particles,
  $\alpha=\A$ (top panel), and the small particles, $\alpha=\B$ (bottom panel),
  of the $\delta=0.60$
  mixture. The full and dashed lines show the MCT plateau values
  $f^c_{\alpha\alpha}(q)=F^c_{\alpha\alpha}(q)/S_{\alpha\alpha}(q)$ calculated
  with the simulation results for the structure factor and the Percus-Yevick
  approximation for the $S_{\alpha\beta}(q)$, respectively.
}
\end{figure}

\begin{figure}
\includegraphics[width=\figwidth]{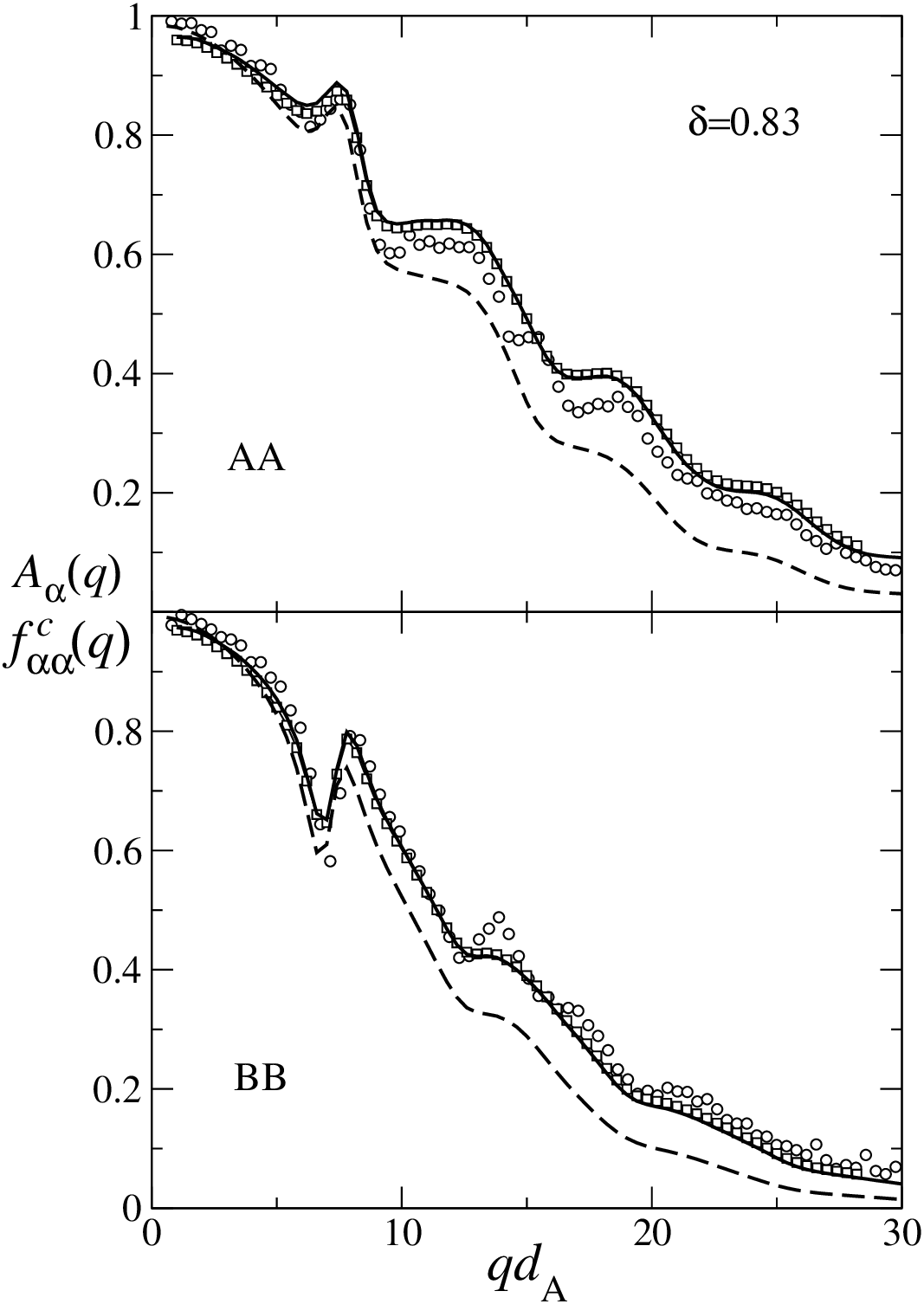}
\caption{\label{mdampl83}
  Analogous results as in Fig.~\protect\ref{mdampl60}, but for the mixture
  with size ratio $\delta=0.83$.
}
\end{figure}

The circles in Figs.~\ref{mdampl60} and \ref{mdampl83} exhibit the
plateau values of the two-step relaxation process, $A_\alpha(q)$, obtained by
fitting
Eq.~\eqref{kww} to the simulation data for the normalized auto-correlation
functions $\phi_\alpha(q,t)$. For $qd_\A<6$, the $A_\alpha(q)$ are very large;
they almost reach their upper limit unity for $q$ tending to zero. This
is a typical mixing phenomenon. For one-component
systems, $A(q)\approx0.4$ is expected for small $q$
\cite{Bengtzelius1984,Fuchs1993}.
The width $q_{1/2}$ of the $A_\alpha(q)$-versus-$q$ curves,
defined by $A_\alpha(q_{1/2})=1/2$, is about $7\%$ larger for $\alpha=\A$ in
Fig.~\ref{mdampl60} than
in Fig.~\ref{mdampl83}. This means that the large particles are better localized
in the mixture with the larger size disparity. For the smaller $\B$ particles,
the opposite trend is observed. Similarly, the plateaus for the
$\A$ correlators exhibit some small peak near the position $qd_\A=7.5$ of the
structure-factor peak, while the $\B$-correlator plateau values exhibit a
pronounced minimum there. For the $\delta=0.60$ mixture, $A_\A(q)$
exhibits a minimum near $qd_\A=11$, which is accompanied by a maximum near
$qd_\A=12$. Instead, the $\delta=0.83$ system has a shoulder for $A_\A(q)$
for $9<q<12$. All these details are reproduced semi-quantitatively by the
results obtained from the MCT values, which are shown as squares in the
figures. There are only a few cases where the plateau values deduced from the
data differ by up to $10\%$ from the ones deduced from the MCT results: the
$\B$ plateaus for the $\delta=0.60$ system for $qd_\A\approx16$ or the
$\A$ plateaus for the $\delta=0.83$ system for $qd_\A\approx17$, for example.

\begin{figure}
\includegraphics[width=\figwidth]{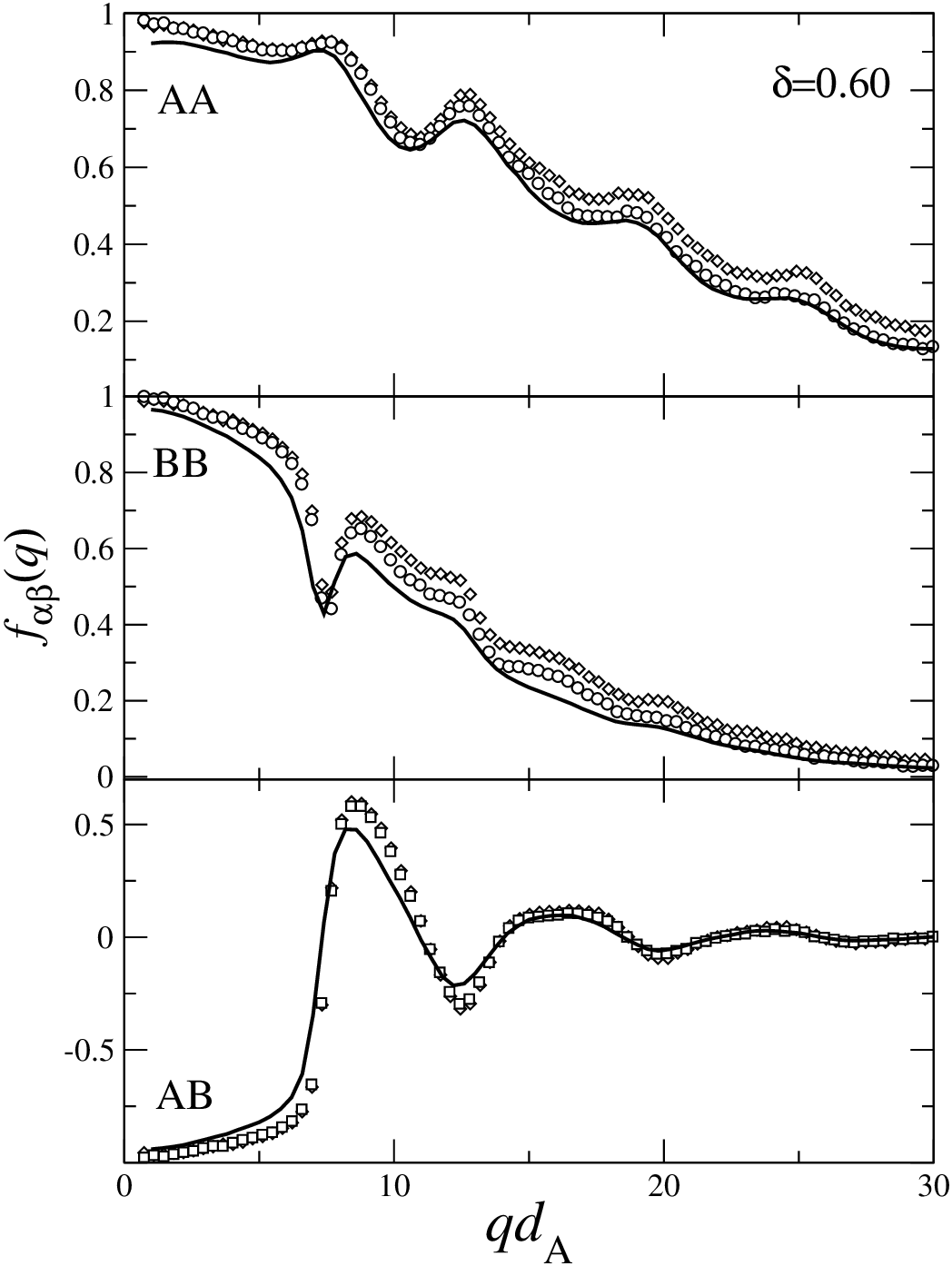}
\caption{\label{vsampl60}
  The circles exhibit the plateau values $f_{\alpha\beta}(q)$ for the
  $\delta=0.60$ mixture, obtained by
  fitting the simulation data for the normalized correlators
  $\Phi_{\alpha\beta}(q)/\sqrt{S_{\alpha\alpha}(q)S_{\beta\beta}(q)}$ to the
  von~Schweidler expansion, Eq.~\protect\eqref{vonschweidler}, see text.
   $\alpha=\A$ refers to the big particles, $\alpha=\B$ to the small ones.
  The fit was optimized to the data for $\varphi=0.59$. The full lines
  are the MCT results for the normalized plateaus $f^c_{\alpha\beta}(q)
  =F^c_{\alpha\beta}(q)/
  \sqrt{S_{\alpha\alpha}(q)S_{\beta\beta}(q)}$. For $\alpha=\beta$, they
  agree with the lines shown in Fig.~\protect\ref{mdampl60}. The squares are
  fit results optimized for the data for $\varphi=0.595$.
}
\end{figure}

\begin{figure}
\includegraphics[width=\figwidth]{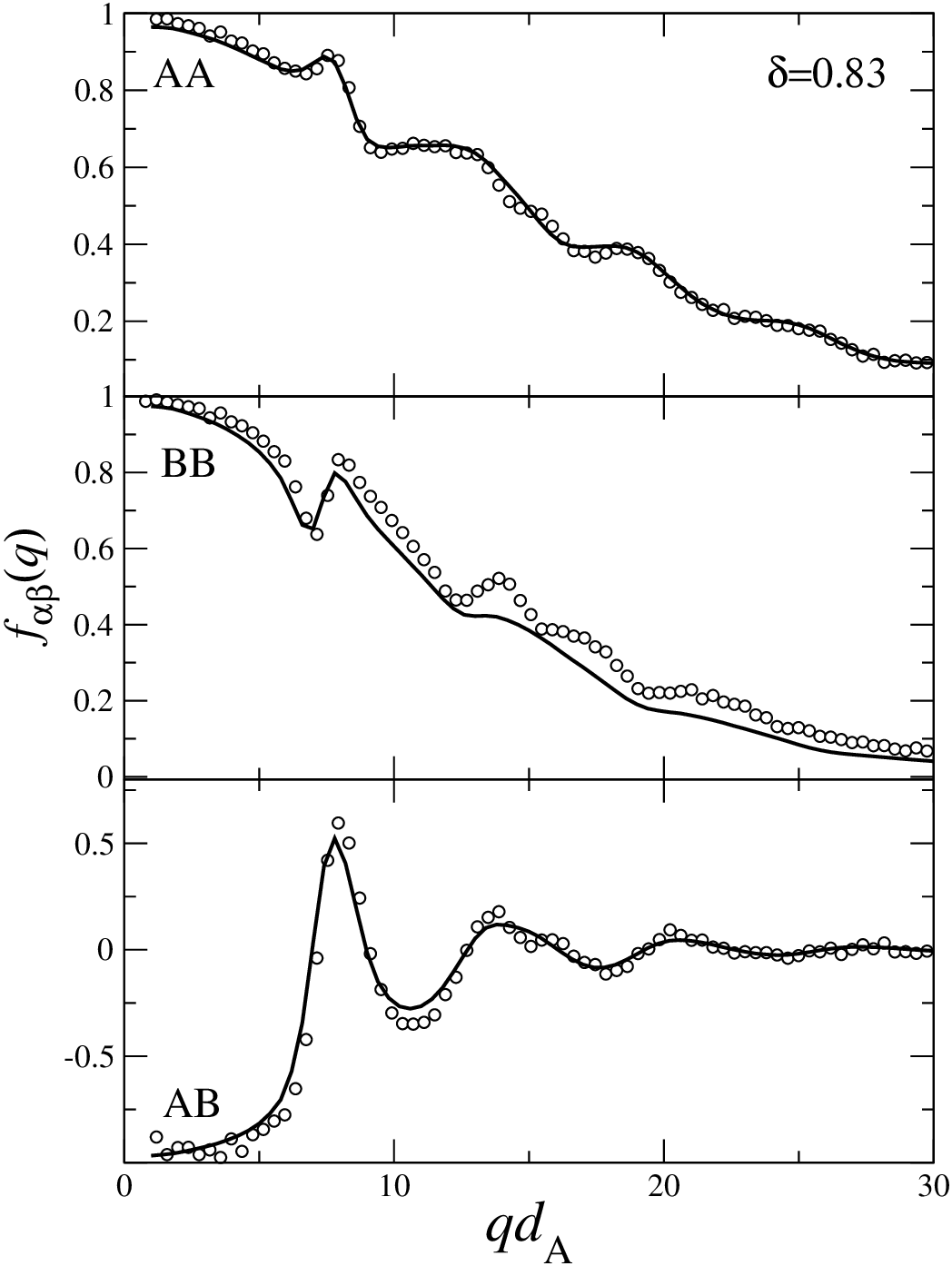}
\caption{\label{vsampl83}
  The analog of Fig.~\protect\ref{vsampl60}, but for the $\delta=0.83$
  system. The fits are based on the data for $\varphi\approx0.582$.
}
\end{figure}

The symbols in Figs.~\ref{vsampl60} and \ref{vsampl83} show the results for
the plateau values $f_{\alpha\beta}(q)$ obtained from fitting
Eq.~\eqref{vonschweidler} to the simulation data, normalized by
$\Phi_{\alpha\beta}(q,t)/\sqrt{S_{\alpha\alpha}(q)S_{\beta\beta}(q)}$.
The fit results obtained for the $\delta=0.60$ mixture for $\varphi=0.590$ and
those for $\varphi=0.595$ differ by about $5\%$. This difference thus
appears as the inherent
uncertainty of the data analysis. No such difference could be identified for
the $\delta=0.83$ mixture. The full lines in the figures exhibit the MCT
plateau values $f^c_{\alpha\beta}(q)=
F^c_{\alpha\beta}(q)/\sqrt{S_{\alpha\alpha}(q)S_{\beta\beta}(q)}$.
For the diagonal functions, $\alpha\beta=\A\A$ and $\alpha\beta=\B\B$,
these lines are identical to the full lines shown in Figs.~\ref{mdampl60}
and \ref{mdampl83}. Those lines should help to compare the plateau results
$A_\alpha(q)$ with the corresponding fit results for the $f_{\alpha\alpha}(q)$.
Obviously, all qualitative features of the plateau fits based on the
Kohlrausch function agree with the ones based on the von~Schweidler series,
both for the fits to the simulation data and for those to the MCT results.
The MCT results for $f^c_{\A\A}(q)$ of the $\delta=0.83$ mixture are in perfect
agreement with the simulation data, while $f_{\B\B}(q)$ is underestimated
systematically by MCT. But the difference is only about $5\%$, except for
$qd_\A\approx14$, where the discrepancy reaches about $10\%$. The deviations
for $f_{\A\B}(q)$ are of similar size. For the system with larger size
disparity, the discrepancies between data and MCT result is somewhat larger,
but it is not seriously larger than the inherent uncertainty of the fits.

Figure~\ref{mctkww} exhibits also the leading term of the von~Schweidler
expansion. In general, accounting for the next-to-leading term of
${\mathcal O}(\tilde t^{2b})$ increases the range of validity of the
von~Schweidler expansion dramatically \cite{Fuchs1993}. Indeed, a data
analysis with a $q$-independent exponent $b$ is possible only if the
${\mathcal O}(\tilde t^{2b})$ term is included \cite{Sciortino1996}.
However, if $\varphi$ is not
close enough to $\varphi^c$, it may happen that the ${\mathcal O}
(\tilde t^{3b})$ terms cancel ${\mathcal O}(\tilde t^{2b})$ contributions.
In such case, the fit range may shrink upon inclusion of the
${\mathcal O}(\tilde t^{2b})$ terms \cite{Franosch1997}. This accident
is demonstrated by Fig.~\ref{mctkww} for $qd_\A=10.2$. Such phenomenon
cannot be foreseen in an unbiased data analysis, which then, necessarily,
must lead to errors in the fit amplitudes. This explains why the sign of the
correction amplitude identified in the lower panel of Fig.~\ref{mdkww}
differs from the one shown in the lower panel of Fig.~\ref{mctkww}.

An obvious source of errors in the MCT results is due to using incorrect
equilibrium-structure information in the mode-coupling functional. It was
mentioned in connection with Eq.~\eqref{fit} that replacing the structure
factors by their Percus-Yevick approximations increases remarkably the
difference between the MCT results for the critical packing fractions
and the results derived from the simulation data.
The dashed lines in Figs.~\ref{mdampl60} and \ref{mdampl83} exemplify the
same phenomenon for the plateau values. These are the MCT results for the
$f^c_{\alpha\alpha}(q)$ based on the use of the Percus-Yevick structure
factors. This approximation for the equilibrium structure leads to
underestimations of the plateau values by more than $10\%$. Figures
\ref{mdampl60} and \ref{mdampl83} demonstrate that MCT is so sensitive to
the small deviations between $S_{\alpha\beta}(q)$ and the Percus-Yevick
results which are exhibited in Fig.~\ref{figsq}, that they lead to serious
flaws in the quantitative predictions for the dynamics.

\begin{figure}
\includegraphics[width=\figwidth]{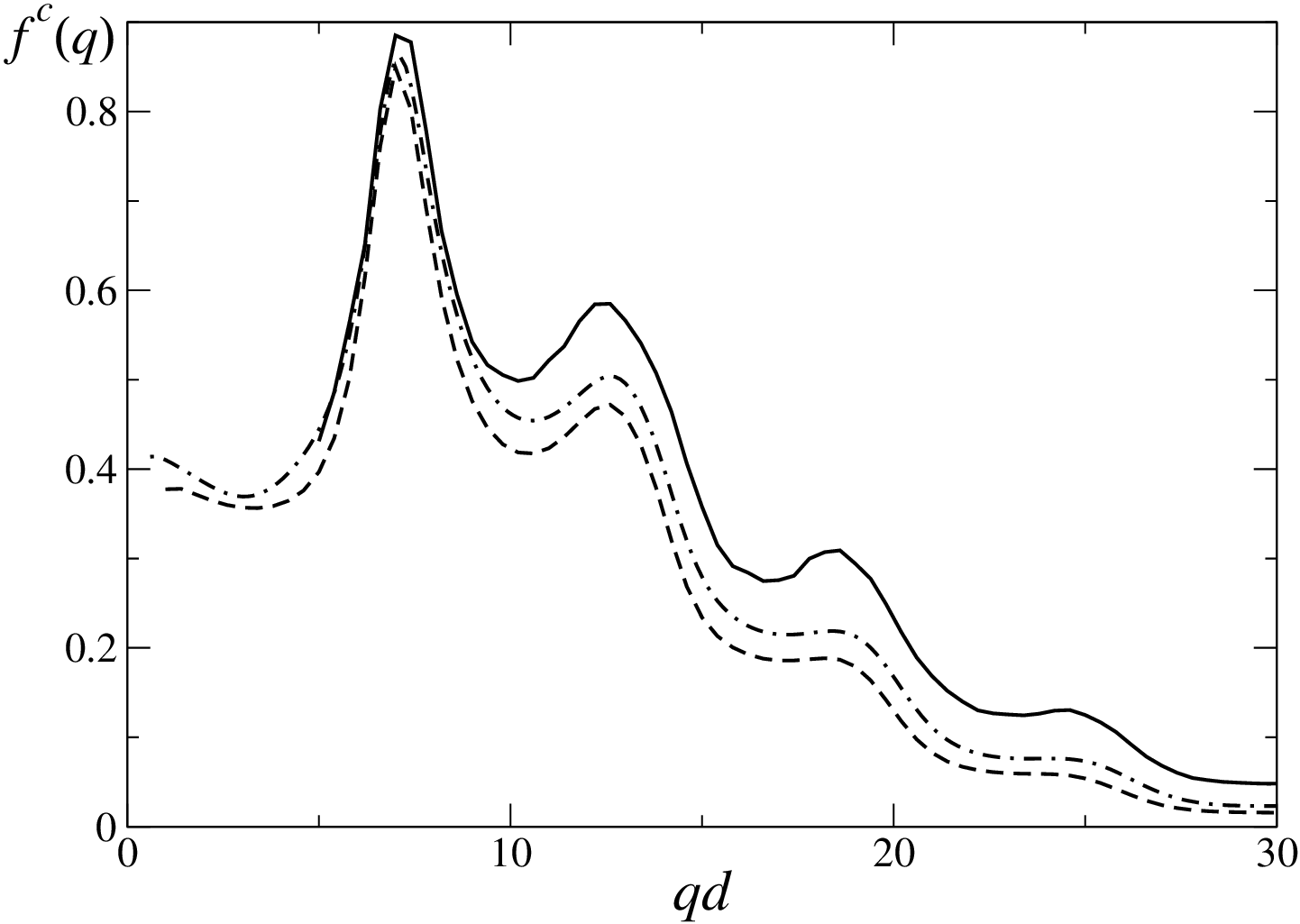}
\caption{\label{fighss}
  MCT plateau values $f^c(q)$ for the one-component hard-sphere
  system, calculated with the simulated structure factor (solid line),
  with the Percus-Yevick approximation (dashed line), and with the
  Verlet-Weis expression for $S(q)$ (dash-dotted line).  The values
  for $qd<5$ based on the simulated structure factor are unreliable
  and have been cut off in the figure.  }
\end{figure}

The difference between the MCT results for the $f^c_{\alpha\beta}(q)$ based
on the Percus-Yevick approximation and those using $S_{\alpha\beta}(q)$ as
obtained from numerical simulations is of similar size for both the
$\delta=0.60$ and the
$\delta=0.83$ system. It is not a specialty of mixtures, as is
demonstrated by Fig.~\ref{fighss}. There, the comparison is repeated for the
one-component hard-sphere system (HSS).
The MCT result for the critical point based on the Percus-Yevick structure
factor is $\varphi^c_{\text{PY}}=0.516$, and the corresponding plateau
values $f^c_{\text{PY}}(q)$ are shown as a dashed line. The Verlet-Weis
modification of the Percus-Yevick structure factor \cite{Verlet1972} is an
empirical improvement of $S(q)$, in particular for the contact values.
Using this as input for the MCT, one gets as the critical point
$\varphi^c_{\text{VW}}=0.525$. The corresponding $f^c_{\text{VW}}(q)$, which
are shown as the dash-dotted line, are systematically larger than the ones
based on the Percus-Yevick results for $S(q)$. We could obtain simulation data
for the structure factor of the metastable HSS for packing fractions up to
$\varphi=0.54$. Beyond this packing fraction, crystallization was always
taking place before particles could diffuse one nearest neighbor distance,
making it impossible to generate data meeting our equilibration criteria.
Extrapolating the smoothly $\varphi$-dependent $S(q)$ up to
$\varphi=0.55$, we calculated as the critical point $\varphi^c_{\text{HSS}}
=0.546$ and obtained the full line as the plateau values. It is remarkable
that the differences between $f^c(q)$ and $f^c_{\text{VW}}(q)$ for
$qd_\A>10$ are larger than those between
$f^c_{\text{VW}}(q)$ and $f^c_{\text{PY}}(q)$.
Note that we find, using the simulation results for the
structure factors, $\varphi^c(\delta\!=\!0.83)<\varphi^c
_{\text{HSS}}<\varphi^c(\delta\!=\!0.60)$, i.e., the system with small
size disparity shows a change of $\varphi^c$ upon mixing that is
qualitatively different from the one seen for large size disparity. This
MCT result is qualitatively the same as predicted originally on the basis
of the Percus-Yevick
approximation \cite{Goetze2003} and is confirmed by our simulations
\cite{Foffi2003}.

\subsection{Von Schweidler-expansion amplitudes}\label{secvs}

\begin{figure}
\includegraphics[width=\figwidth]{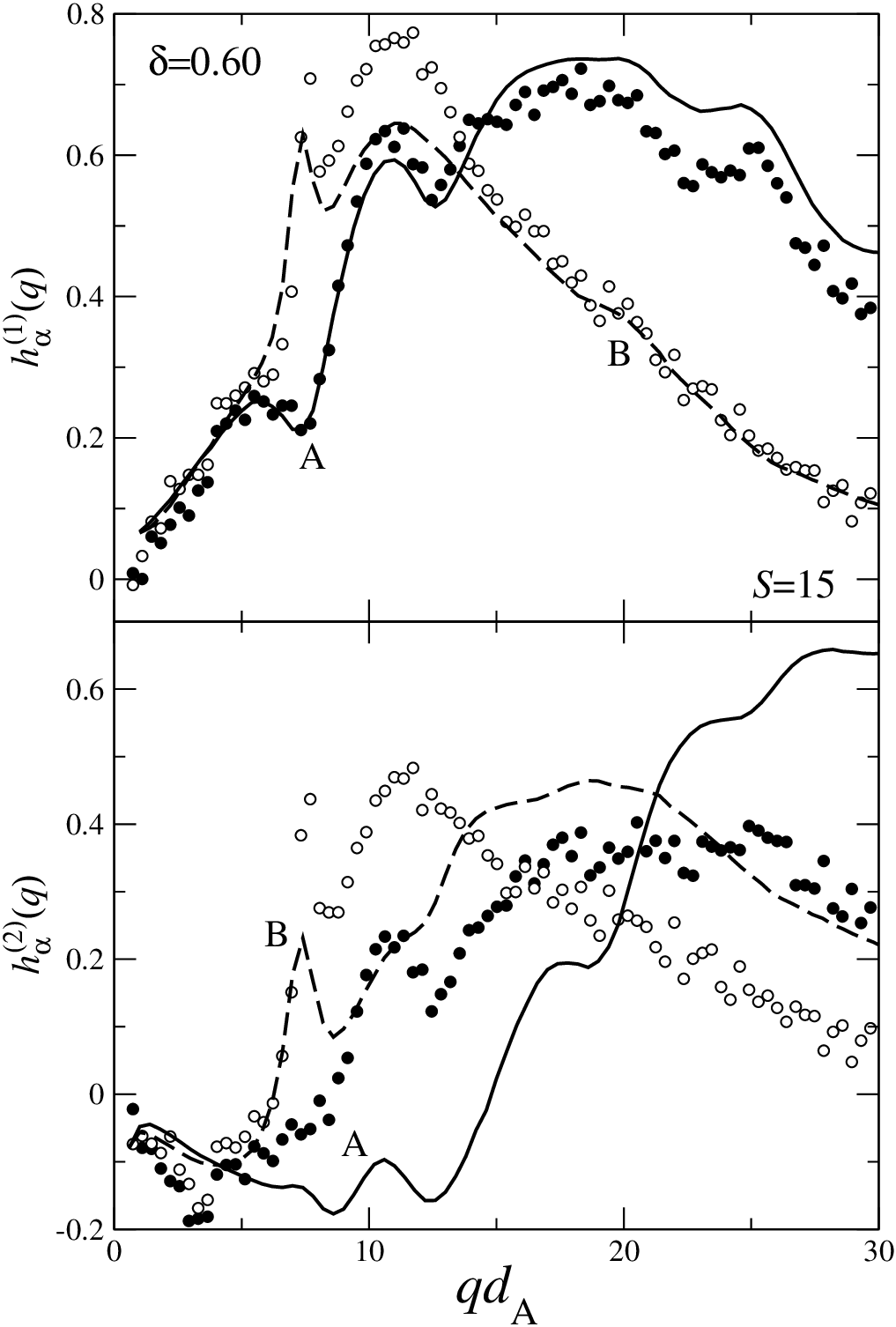}
\caption{\label{figh}
  Amplitudes
  $h_\alpha^{(1,2)}(q)=H_{\alpha\alpha}^{(1,2)}(q)/S_{\alpha\alpha}
  (q)$ for the von~Schweidler-law expansion of the normalized density
  auto-correlation functions $\Psi_{\alpha\alpha}(q,\tilde t)
  /S_{\alpha\alpha}(q)$ for the $\delta=0.60$ mixture. The dots and
  circles are the results of fitting Eq.~\protect\eqref{vonschweidler}
  to the simulation data for $\alpha=\A$ (large) and $\alpha=\B$
  (small particles), respectively. The full and dashed lines are the
  corresponding results calculated from the MCT equations. Scale
  factors $S=15$ and $S^2$ are used for the comparisons of the
  $h_\alpha^{(1)}(q)$ and $h_\alpha^{(2)}(q)$, respectively, to
  account for the different time scales $\tau$ used in the analysis of
  the simulation data and of the calculations.  }
\end{figure}

The amplitudes $H_{\alpha\beta}^{(1)}(q)$ in Eq.~\eqref{vonschweidler}
are the most important parameters quantifying the dynamics in a time
interval where the correlators are close to their plateaus. The upper panel
of Fig.~\ref{figh} exhibits a set of representative results. It compares
the amplitudes $h_\alpha^{(1)}(q)$ of the normalized master functions
$\Psi_{\alpha\alpha}(q,\tilde t)/S_{\alpha\alpha}(q)$ obtained from the
analysis of the simulation data for the $\delta=0.60$ mixture with the
corresponding quantities calculated within MCT. For the quantitative
comparison, a scale factor $S$ has to be adjusted since the arbitrariness of the
time scale $\tau$ implies an arbitrariness in the prefactor $\tau^{-b}$ of
$h_\alpha^{(1)}(q)=H_{\alpha\alpha}^{(1)}(q)/S_{\alpha\alpha}(q)$.
The $h_\alpha^{(1)}$-versus-$q$ curves exhibit a subtle structure. While
$h_\A^{(1)}(q)$ has a maximum for $qd_\A\approx5.7$ followed by a
minimum for $qd_\A\approx7.4$, i.e.\ for a wave vector near the structure-factor
peak position, $h_\B^{(1)}(q)$ increases monotonically to a maximum for
$qd_\A\approx7.4$. While $h_\A^{(1)}(q)$ increases for $qd_\A>7.4$ monotonically
to a maximum for $qd_\A\approx11$, $h_\B^{(1)}(q)$ exhibits a sharp minimum
for $q$ slightly above $7.4$
before it also reaches a maximum for $qd_\A\approx11$. For $qd_\A>11$,
$h_\B^{(1)}(q)$ decreases monotonically, while $h_\A^{(1)}(q)$ has a minimum
for $qd_\A\approx12.7$ and then exhibits a broad maximum. These features
are reproduced by the MCT results. The MCT results agree with the data on a
$10\%$ level, except for the $\B$ amplitudes for $qd_\A$ near $10$, where there
are $20\%$ discrepancies.

The amplitudes $h_\alpha^{(2)}(q)$, which describe corrections to
von~Schweidler's law $\Psi_{\alpha\alpha}(q,\tilde t)/S_{\alpha\alpha}(q)
=f^c_{\alpha\alpha}(q)-h_\alpha^{(1)}(q)\tilde t^b$, exhibit a zero at some
wave vector $q_\alpha^*$. For $q<q_\alpha^*$, the amplitudes are negative,
and for $q>q_\alpha^*$ they are positive. These features and also the value
$q_\B^*\approx6$ are reproduced by the MCT results.
Notably, the MCT results for $qd_\A\le12$ still share some qualitative features
with the results obtained from the fit to the simulation data, e.g.\ the
sharp peak followed by a sharp minimum in $h_\B^{(2)}(q)$ at $qd_\A\approx7$,
and the peak in $h_\A^{(2)}(q)$ at $qd_\A\approx11$.
Otherwise, one notices
serious discrepancies between data and MCT results. For example, MCT predicts
a particularly large range of validity of von~Schweidler's law for the
density fluctuations of the large particles for a wave vector
$qd_\A\approx15$. But the data analysis is done best for this case by using a
correction amplitude near $0.2$.
The identified discrepancies signalize the limitations in the
determination of a correction amplitude for an asymptotic law from data
which cannot be chosen sufficiently close to the singularity.

\subsection{The stretching exponents}

\begin{figure}
\includegraphics[width=\figwidth]{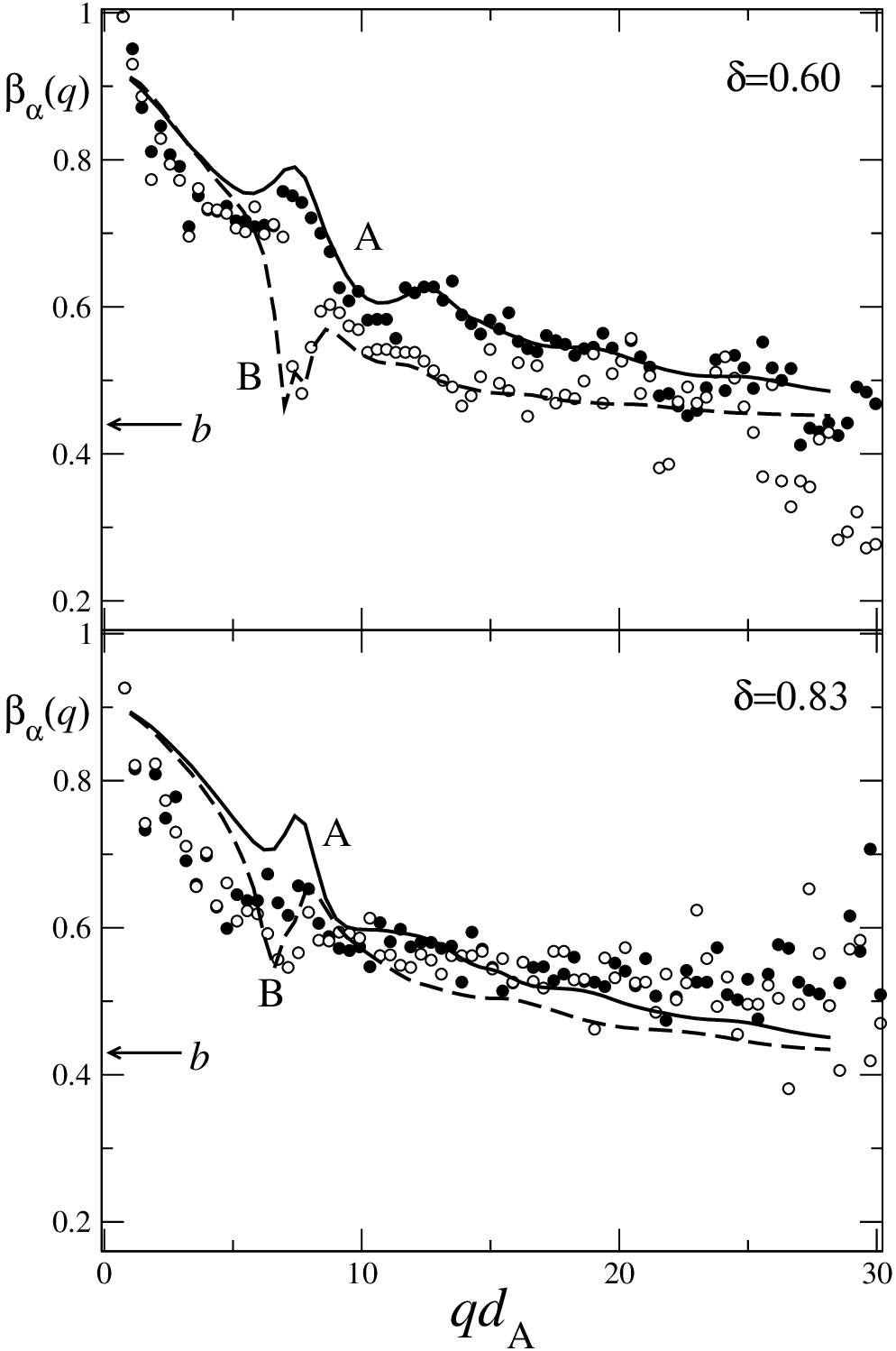}
\caption{\label{figbeta}
  The stretching exponents $\beta_\alpha(q)$ obtained by fitting the
  Kohlrausch function, Eq.~\protect\eqref{kww}, to the alpha-relaxation
  master functions for the density auto-correlation functions of the two
  mixtures with size ratios $\delta$ of the particles $0.60$ and $0.83$.
  The arrows in the two panels mark the von~Schweidler exponents $b$,
  Eqs.~\protect\eqref{fit}. The dots and circles are obtained by fits to
  the simulation data for the big particles, $\alpha=\A$, and the small ones,
  $\alpha=\B$, respectively. The full and dashed lines are the corresponding
  results obtained by fits to the solutions of the MCT equations.
}
\end{figure}

The exponent $\beta$ in Eq.~\eqref{kww} provides a convenient overall measure
for the alpha-relaxation stretching. It quantifies in an averaged manner
deviations of the alpha-relaxation process from a Debye law,
$\phi(t)\propto\exp(-\tilde t/\tilde\tau)$. The latter is the universal result
for the
dynamics of a variable coupled to a white-noise field. Figure~\ref{figbeta}
shows that for $2<q<6$, the exponent decreases considerably with increasing
$q$. There is no difference between the fluctuations for the large and the
small particles in this wave-vector interval; and the stretching is larger
for the system with smaller size disparity. For the wave vectors near the
structure-factor-peak position, $qd_\A\approx7$, the stretching of the 
$\B$-fluctuations of the $\delta=0.60$ mixture is much bigger than the one
of the $\A$-fluctuations: $\beta_\B\approx0.5$ versus $\beta_\A\approx0.75$.
There is an indication of the same phenomenon for the $\delta=0.83$ mixture.
For larger wave vectors, $\beta_\A$ is somewhat larger than $\beta_\B$ for the
mixture with large size disparity. For the $\delta=0.83$ mixture, $\beta_\A$
equals $\beta_\B$ for $qd_\A>8$ within the noise of the data.

For $qd_\A<6$, the MCT results overestimate $\beta_\alpha(q)$ by about $10\%$. The
$qd_\A\approx7$ anomaly and also the large-$q$ variation for the $\delta=0.60$
mixture are described well by the theory. For the system with small size
disparity, MCT overestimates the $qd_\A\approx7$ anomaly, and there is a slight
trend to underestimate $\beta_\alpha(q)$ for large $q$.

For large wave vectors, the MCT alpha-process-master functions approach the
Kohlrausch law. In this asymptotic regime, $\beta_\alpha(q)=b$
\cite{Fuchs1994}. Figure~\ref{figbeta} illustrates that this
theoretical result is consistent with the simulation data.

\subsection{The alpha-relaxation time scale}\label{secalphascale}

\begin{figure}
\includegraphics[width=\figwidth]{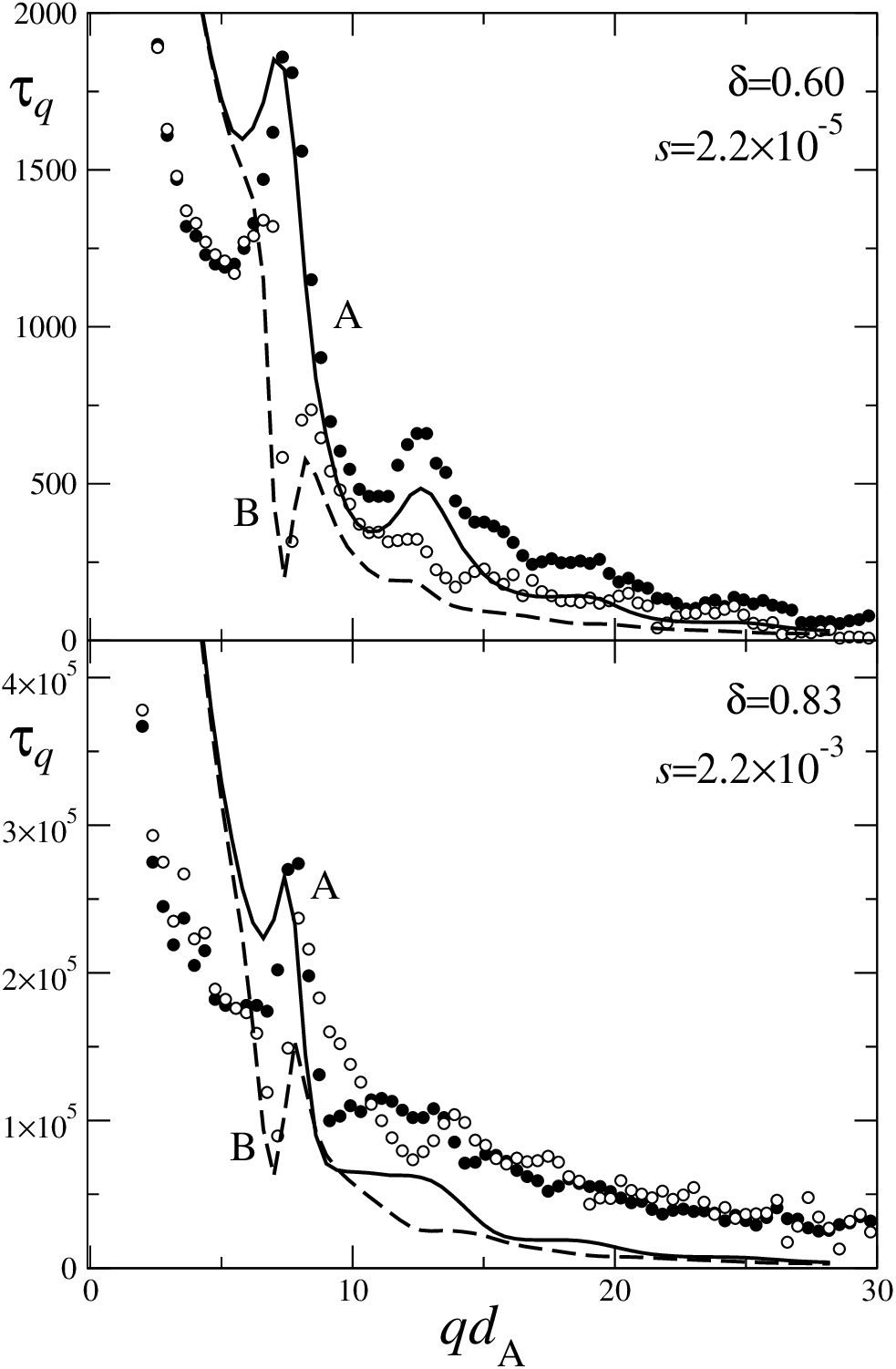}
\caption{\label{figtauq}
  Alpha-relaxation time scales $\tau_\alpha(q)$ obtained by fitting
  Eq.~\protect\eqref{kww} to the data (symbols) and MCT results (lines).
  The coding is the same as used in Figs.~\protect\ref{figh} and
  \protect\ref{figbeta}. A
  multiplication factor $s$ is applied to the MCT results in order to
  account for the different time scales relative to that used for the data.
}
\end{figure}

The fit of Eq.~\eqref{kww} to the long-time parts of the correlators yields
the time scales $\tilde\tau_\alpha(q)$ for the alpha processes up to an overall
scale $\tau$. The results for the fits allow for a comparison of the scales for
the fluctuations of the large particles with those for the small ones. They
also allow to discuss the $q$ dependence of the relaxation times. In order to
compare the scales from the data with those calculated from MCT, one has to
fit an overall scale factor $s$.

Figure~\ref{figtauq} exhibits the results for the two mixtures.
For $qd_\A\le6$, $\tilde\tau_\A(q)=\tilde\tau_\B(q)$,
and both scales decrease with increasing
wave vector $q$. These features are reproduced by MCT, but the theory
overestimates the time scales seriously. For wave vectors near the 
structure-factor-peak position, $\tilde\tau_\A$ exhibits a pronounced maximum,
and $\tilde\tau_\B$ has a sharp minimum.
The ratios of the scales is about $5$ and
$3$ for the systems with large and small size disparity, respectively. This
feature is reproduced well by MCT. For the $\delta=0.60$ mixture,
$\tilde\tau_\A(q)$ exhibits a maximum for $qd_\A\approx13$, while
$\tilde\tau_\B(q)$ has a
shoulder there. For $qd_\A\ge15$, the relaxation times decrease with increasing
$q$. The time scales for the large particles are somewhat larger than those
for the small particles. These features are reproduced qualitatively by MCT,
but the theory underestimates the time scales by a factor $2$ to $3$ for
wave vectors above the structure-factor-peak position. For the
$\delta=0.83$ mixture, the scale $\tilde\tau_\A(q)$ exhibits a shoulder for
$10<q<13$ in accord with MCT. The time $\tilde\tau_\B(q)$ exhibits a minimum for
$qd_\A\approx12$, while MCT shows a kink there. Again, MCT underestimates the
$\tilde\tau_\alpha(q)$ for $qd_\A\ge10$.

\begin{figure}
\includegraphics[width=\figwidth]{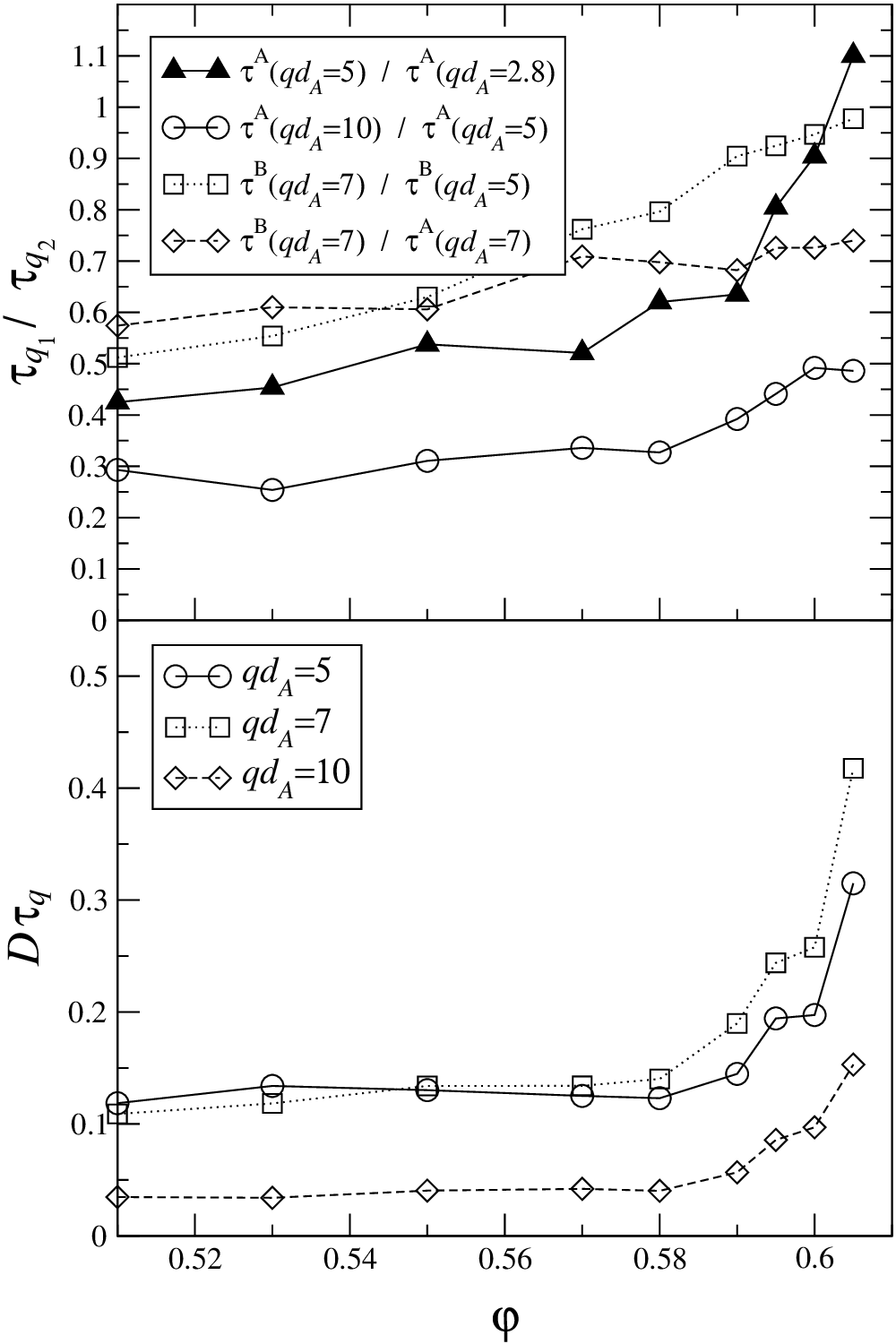}
\caption{\label{figdtau}
  Ratio of alpha-relaxation times $\tau^\alpha(q)$ (upper panel) and product
  of alpha-relaxation times with the  large-particle diffusivity $D=D_\A$
  (lower
  panel) for the $\delta=0.60$ system as functions of the packing fraction
  $\varphi$ (see text). The lines are guides to the eye.
}
\end{figure}

Let us consider the variation of the alpha-relaxation time scales as
functions of the packing fraction $\varphi$. To this end, we have determined
a time scale $\tau^\alpha(q,\varphi)$ for this process by arbitrarily choosing
$\phi_{\alpha\alpha}(q,t\!=\!\tau^\alpha(q,\varphi))=0.1$, for those values
of $q$ where the plateau values are still appreciably larger than $0.1$.
Equation \eqref{alphascaling}
formulates the scale coupling of the MCT results. While the time scales
of two alpha processes, say $\tau_1$ and $\tau_2$, diverge for vanishing
distance parameter $\varepsilon=(\varphi^c-\varphi)/\varphi^c$, the ratio
$\tau_1/\tau_2$ is a smooth function of $\varepsilon$. For example, let
$\tau_1$ and $\tau_2$ refer to the alpha processes of the correlator
$\phi_{\A\A}(q,t)$ of the $\delta=0.60$ system for $qd_\A=10$ and $qd_\A=5$,
respectively. If $\varepsilon$ increases up to about $0.05$,
$\tau_1/\tau_2$ decreases linearly with $\varepsilon$ by about
$4\%$. The simulation results behave similarly, as is shown in the
upper panel of Fig.~\ref{figdtau}.
If the packing fraction
increases from $0.50$ to $0.605$, the alpha-relaxation time scales of the
density-fluctuation correlators increase by more than three orders of
magnitude. Still, the shown three representative ratios for the scales vary by
less than a factor two.
Hence, the scale-coupling prediction is verified on a $10\%$ level for the
three ratios shown in the upper panel of Fig.~\ref{figdtau} by the open
symbols. These examples are representative for density-fluctuation scales
with intermediate and large wave numbers. If one of the wave numbers
decreases to small values, the violation of the scale coupling becomes larger,
as is demonstrated by the full symbols in Fig.~\ref{figdtau}.
The diffusivity $D$ is proportional to the inverse of the alpha-relaxation
scale $\tau_D$ of the mean-squared displacement. Hence, a
$\tau\cdot D$-versus-$\varphi$ diagram demonstrates the coupling of the
scales for the processes described by $\tau$ and that for the diffusivity.
The lower panel of Fig.~\ref{figdtau} shows the results for the
$\delta=0.60$ system. Here, $\tau$ refers to the scale for the density
fluctuations of the $\A$ particles, and the $D$ are the simulation data
for the $\A$-particle diffusivities \cite{Foffi2003}. The scale coupling holds
for $\varepsilon\ge0.05$. However, approaching the transition point more
closely, contrary to the MCT results, the scale for the diffusivity decouples
from that for the density fluctuations. The diffusivity does not decrease
with decreasing $\varepsilon$ as strongly as $1/\tau$. The results of our
simulations for the $\delta=0.83$ system behave similarly.
The described decoupling is in qualitative agreement with the behavior
found earlier for the simulation results of a binary Lennard-Jones
system \cite{Kob1994} and for a model for water
\cite{Starr1999c}.

\begin{figure}
\includegraphics[width=\figwidth]{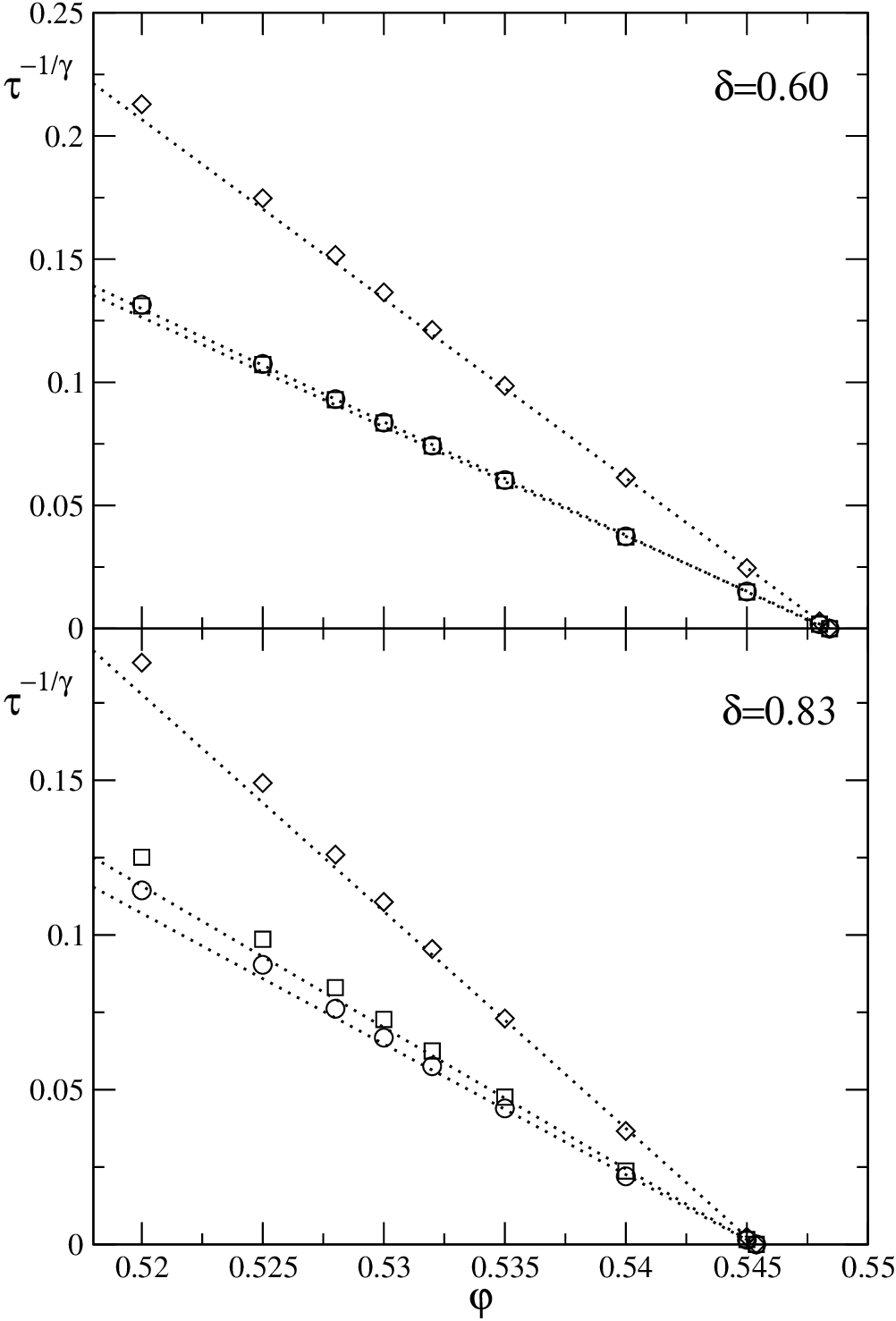}
\caption{\label{figrect}
  $\tau^{-1/\gamma}$ with $\gamma=3$ as function of the packing
  fraction $\varphi$. The $\tau$ are the alpha-relaxation times for
  the large-particle density-fluctuation correlators
  $\phi_{\A\A}(q,t)$ calculated from the MCT equations, defined by
  $\phi_{\A\A}(q,\tau)=0.1$. The wave numbers $q$ are $10.2$
  (diamonds), $7.4$ (squares), and $5.0$ (circles). The dotted lines
  are the asymptotic laws $\tau^{-1/\gamma}\propto(\varphi^c-\varphi)$
  with $\varphi^c(\delta=0.60)=0.548$ and $\varphi^c(\delta=0.83)
  =0.545$.  }
\end{figure}

\begin{figure}
\includegraphics[width=\figwidth]{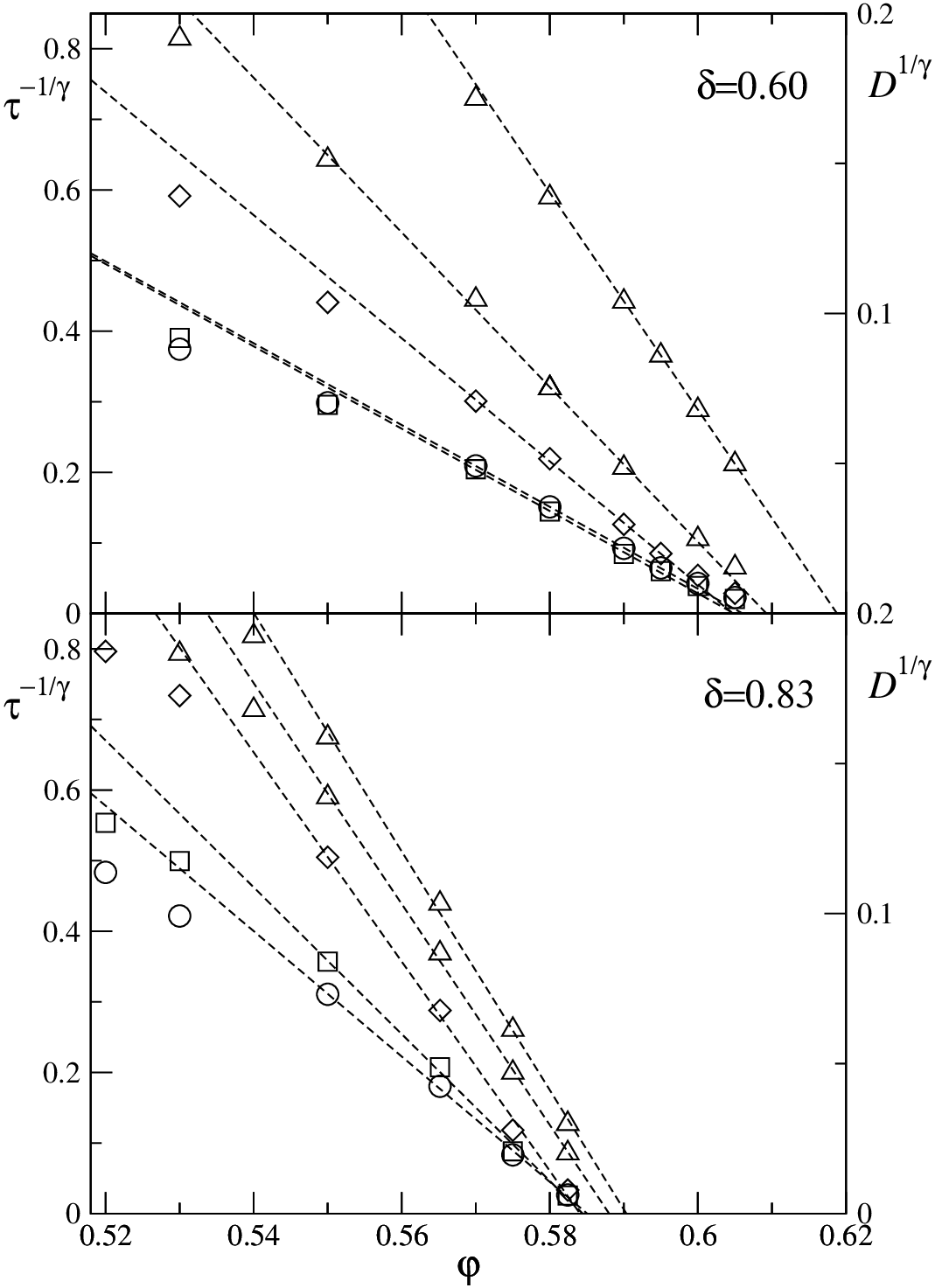}
\caption{\label{mdrect}
  $\tau^{-1/\gamma}$ with $\gamma=3$ (left scale) as function of
  $\varphi$.  The $\tau$ are the alpha-relaxation times for the
  large-particle correlation functions at wave number $qd_\A=10$
  (diamonds), $7$ (squares), and $5$ (circles), defined by
  $\phi_{\A\A}(q,\tau)=0.1$.  The upper and lower triangles are
  $D^{1/\gamma}$ (right scale) for the diffusivities of the small and
  big particles, respectively. The dashed lines are linear
  interpolations of the data for large $\varphi$.  }
\end{figure}

MCT predicts a power-law divergence in the asymptotic limit of vanishing
$\varepsilon$ for the common scale $\tau$ in Eq.~\eqref{alphascaling}:
$\tau\propto\varepsilon^{-\gamma}$. The exponent $\gamma$ is determined by
the von~Schweidler exponent $b$ \cite{Goetze1992}. The value $b\approx0.44$
used throughout the preceding discussions implies $\gamma\approx3.0$.
Figure~\ref{figrect} demonstrates this property for the MCT results for
$\tau_\A(q)$ for three representative wave numbers $q$ in form of a
rectification diagram. In agreement with typical results for the simple
HSS \cite{Franosch1997}, the asymptotic description holds well for
$\varepsilon$ up to about $0.05$, and there appear deviations if $\varphi$
differs from $\varphi^c$ by more than $5\%$. Analogous rectification
diagrams for the simulation data are shown in Fig.~\ref{mdrect}.
Linear extrapolations of the data to large $\varphi$ for three wave numbers
$q$ yield estimates for the critical packing fractions:
$\varphi^c(\delta=0.60)=0.606\pm0.001$ and $\varphi^c(\delta=0.83)=0.585\pm
0.001$.
The rectification curves for the diffusivity relaxation times for both types
of particles are included in the figure as $D^{1/\gamma}$-versus-$\varphi$
plots. The data seem to follow the power-law predictions, but lead to
slightly different estimates of $\varphi^c$, also depending on the species
$\alpha=\A,\B$. For the $\delta=0.60$ system, we get
$\varphi^c_{D_\A}\approx0.609$ and $\varphi^c_{D_\B}\approx0.619$, while for
the $\delta=0.83$ system,
$\varphi^c_{D_\A}\approx0.588$ and $\varphi^c_{D_\B}\approx0.590$.
The decoupling of the diffusivity scales from the ones for the density
fluctuations mentioned above yields this overestimation of the critical
packing fractions.
Let us emphasize that the described
estimations of $\varphi^c$ have been done with the bias of a given exponent
$\gamma=3.0$. An unbiased three-parameter fit of the scale as a function of
$\varphi$ by the formula $\tau=C(\varphi^c-\varphi)^{-\gamma}$ suffers from
correlations between the fit parameters $\varphi^c$ and $\gamma$.
Indeed, such fits to the diffusivities of the two species lead to differing
exponents $\gamma$ \cite{Foffi2003}, in disagreement with MCT. A similar result
has also been found in a simulation of a binary Lennard-Jones mixture
\cite{Kob1994}.
Fig.~\ref{mdrect} however shows that on the basis of our
simulation data, one can, given the restricted range of validity of the
asymptotic law, not distinguish between these different $\gamma$ values.
Note that the largest discrepancy, either in $\gamma$ or in $\varphi^c$,
emerges for the $\B$ particles in the $\delta=0.60$ system.
Unbiased fits lead to larger uncertainties for $\varphi^c$, since a decrease
of  the fit parameter $\varphi^c$ can partly be compensated by a decrease
of  the fit parameter $\gamma$. One could
get the crossing points of the $D^{1/\gamma'}$-versus-$\varphi$ curves
closer to that of the $\tau^{-1/\gamma}$-versus-$\varphi$ curves in
Fig.~\ref{mdrect}, if one would use some $\gamma'<\gamma$. Such formulation
of the decoupling phenomenon is suggested by Fig.~\ref{figdtau},
since the increase in $D\cdot\tau$ for $\varphi$ increasing above $0.58$
can be fitted by $(\varphi^c-\varphi)^x$, $x=\gamma'-\gamma<0$.

A comparison of Figs.~\ref{figrect} and \ref{mdrect} leads to two questions,
which we cannot answer. Why is the range of the distances $\varepsilon$,
where the power-law asymptote describes the alpha scales, so much larger
for the simulation results than for the MCT solutions? Why do the deviations
of $\tau^{-1/\gamma}$ for larger $\varepsilon$ from the small-$\varepsilon$
asymptote have a different sign for the simulation results than for the
MCT solutions?

\section{The alpha-process shape functions}\label{sec.amaster}

\begin{figure}
\includegraphics[width=\figwidth]{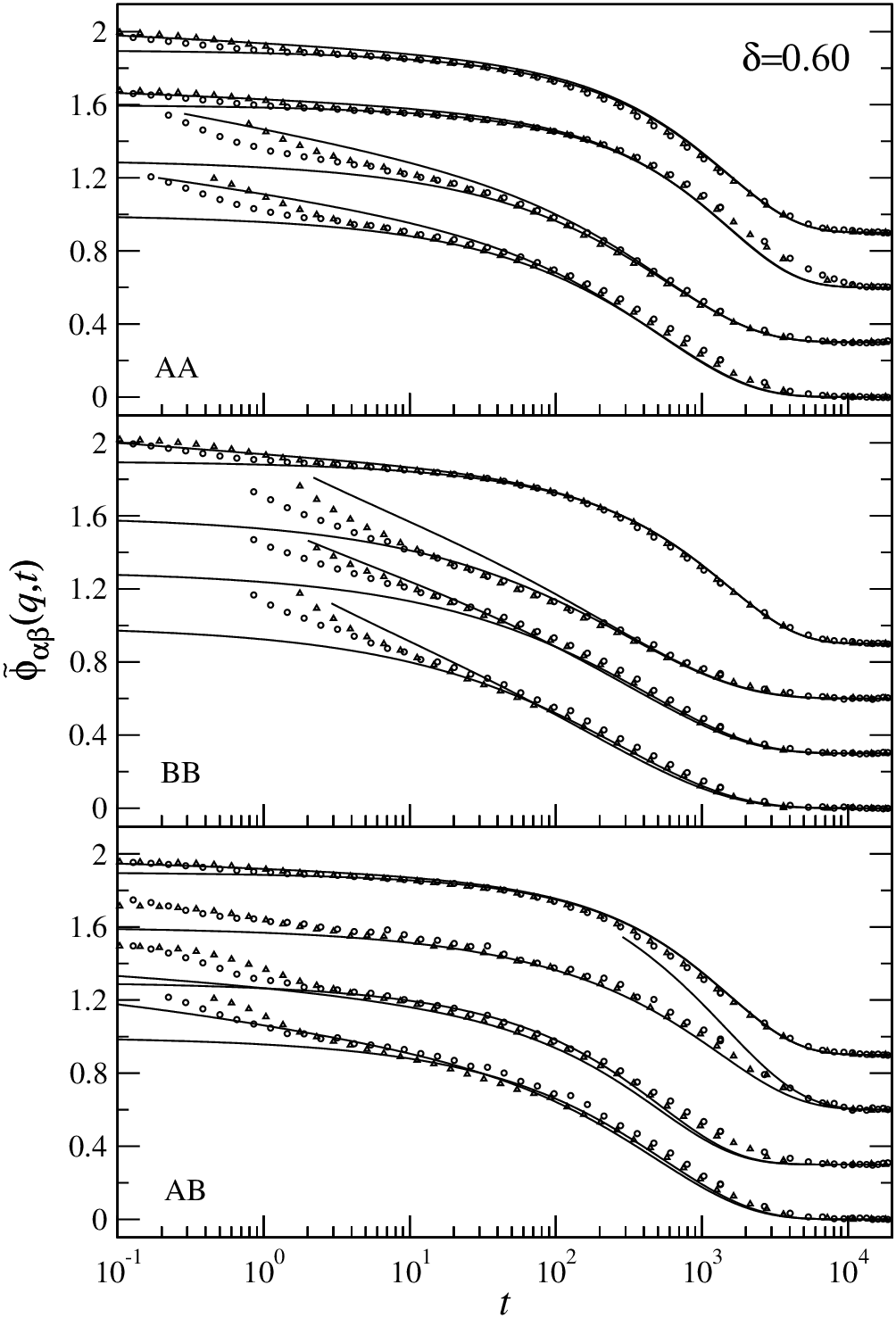}
\caption{\label{alpha-master60}
  Alpha-master functions
  $\tilde\phi_{\alpha\beta}(q,t)=\phi_{\alpha\beta}(q,t)
  /f_{\alpha\beta}(q)$ for the $\delta=0.60$ mixture ($\alpha=\A$
  refers to large particles, $\alpha=\B$ to small ones). The circles
  and triangles represent the simulation data for $\varphi=0.595$ and
  $0.59$, respectively, scaled as in Fig.~\protect\ref{alphasc}. The
  solid lines are MCT results for the master function
  $\Psi_{\alpha\beta}(q,t)/F^c_{\alpha\beta}(q)$, and a solution for a
  distance $\varepsilon=(\varphi-\varphi^c)/\varphi^c \approx-0.015$
  from the critical point. Time has been rescaled for all curves in
  order to match the alpha-time scale of the simulation data at $qd_\A=5$,
  $\alpha\beta=\A\A$.  The results refer to wave vectors $qd_\A=5$, $7.5$,
  $10$, and $12.5$ (from top to bottom), where the curves for
  different $q$ have been shifted vertically by $0.3$ for enhanced
  clarity.  }
\end{figure}

\begin{figure}
\includegraphics[width=\figwidth]{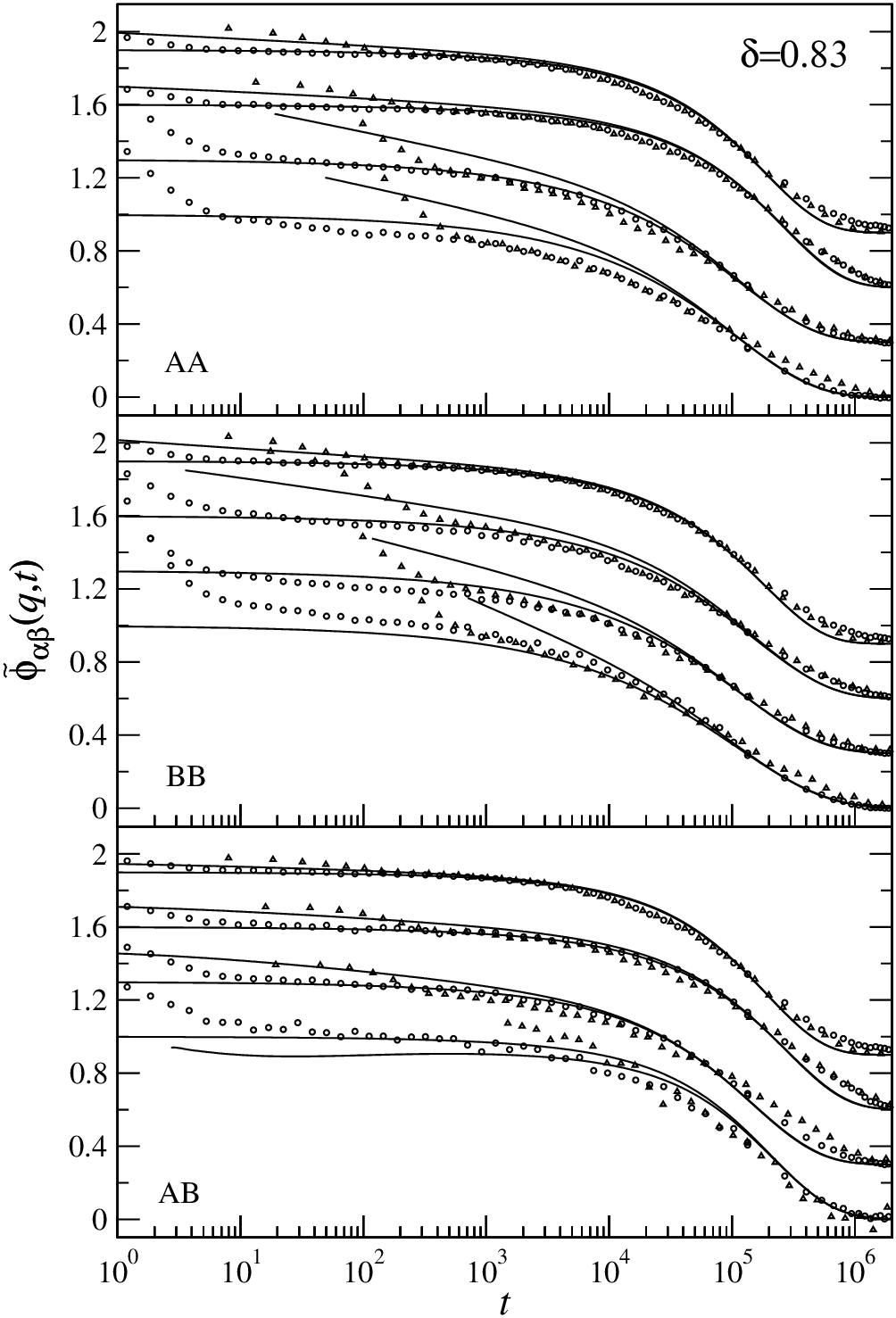}
\caption{\label{alpha-master83}
  As in Fig.~\protect\ref{alpha-master60}, but for $\delta=0.83$ and
  $\varphi\approx0.582$ (circles) and $\varphi\approx0.575$ (triangles).
  The solid lines are the theoretical master curve and a curve with
  $\varepsilon\approx-0.01$.
}
\end{figure}

The preceding parameterization of the correlation functions shows that there
is no universal master function for the alpha-relaxation processes. For
example, the stretching of the density-fluctuation auto-correlation functions
for the large particles generally differs from the one for the small particles,
and it depends on the wave vector of the fluctuations. It is a challenge for a
microscopic theory to describe the alpha-process shape functions for different
probing variables. Figures~\ref{alpha-master60} and
\ref{alpha-master83} present our simulation
results for the two mixtures under consideration for four representative wave
vectors in comparison with the MCT curves. All results are rescaled by the
plateau values and by the time scales discussed in Sec.~\ref{sec.param}.
The data are shown for two packing fractions in order to document the
asymptotic regime of validity of the superposition principle. The MCT master
function is complemented by a curve for some small $\varphi^c-\varphi$,
in order to indicate the effect of preasymptotic corrections to the
alpha-process asymptote.

Figure~\ref{alpha-master60}
shows that, typically, the decay of the correlators
of the $\delta=0.60$ system from $90\%$ to $5\%$ of their respective plateau
values is stretched on a time interval of about three decades. With two
exceptions, this decay is described well by the MCT master functions for the
alpha process. The first exception is the $\A\A$ correlator for
$qd_\A=7.5$ for rescaled times $t\approx 4\times10^3$. Here, the data decay
less rapidly as the exponential long-time tail obtained by MCT. The second
exception concerns the $\B\B$ correlator for $qd_\A=10$ and $qd_\A=12.5$. Here,
the data fall on the master functions only for such long times, where the
correlator is below $70\%$ of the plateau value. But, the MCT results for
$(\varphi-\varphi^c)/\varphi^c\approx-0.015$, which are added in rescaled
presentation as full lines in Fig.~\ref{alpha-master60},
exhibit the same phenomenon.
The reason for this is the large size of the critical amplitude
$h_\B^{(1)}(q)$ for these wave vectors, cf.\ Fig.~\ref{figh}.
They cause particularly large preasymptotic corrections to the alpha-scaling
law near the plateau. Thus, the specified exceptions are not a defect of
MCT, but a confirmation of a subtlety of that theory.

The test of the alpha-process shape functions for $\delta=0.83$,
Fig.~\ref{alpha-master83},
exhibits a series of problems. Deviations of the data
from exponential decay for the very long rescaled times $t\approx5\times
10^5$ occur for $qd_\A=5$ and $qd_\A=7.5$ for all three correlators. The
$\A\B$ correlator for $qd_\A=12.5$ shows more stretching than the MCT solutions.
Furthermore, the $\B\B$ correlators for $qd_\A=10$ and $qd_\A=12.5$ miss the
plateau. However, the latter is an obvious mistake of the data analysis, which
could be eliminated by correcting the plateau value. We did not carry out
this correction in order to emphasize that $5\%$ errors in the determination
of the plateau values are almost unavoidable in an unbiased data analysis.
The reduction of the scaling regime for the $\B\B$ correlator for $qd_\A=10$
and $12.5$ occurs for the $\delta=0.83$ system as discussed above for the
$\delta=0.60$ system; and it can be explained in the same manner.

The most severe problem exhibited by Fig.~\ref{alpha-master83}
is the following.
Even for $\varepsilon\approx0.02$, in the simulation, several
correlators stay within a $5\%$ interval around the plateau for time
intervals as large as $2.5$ decades or more. This holds, e.g., for the
$\A\A$ correlator for $qd_\A=7.5$, for the $\B\B$ correlator for $qd_\A=5$, and for
the $\A\B$ correlator for $qd_\A=7.5$ and $10$. The full lines show that this
feature cannot be explained by MCT. Even if the distance parameter
$\varepsilon=(\varphi^c-\varphi)/\varphi^c$ is as small as $10^{-2}$, the
calculated correlators cross their plateaus much steeper than exhibited by
the simulation data.

\section{Conclusions}\label{sec.conc}

Molecular-dynamics simulations have been presented for two dense binary
mixtures of hard spheres. One mixture deals with a size ratio $\delta=0.60$
for the two particle species and the other with $\delta=0.83$. The first
system is representative for the situation where mixing stabilizes the
liquid state, and the other for the one where mixing stabilizes the glass
\cite{Foffi2003}. The data demonstrate the evolution of a two-step relaxation
scenario with increasing packing fraction $\varphi$,
Figs.~\ref{figphia} and \ref{figphib}, which is similar to the one detected
previously for other systems. In this paper, a comprehensive analysis of
the second relaxation step, usually referred to as the alpha process, is
presented. It deals with the decay of the correlators from some
plateau to zero. The process was identified as that part of the correlators
exhibiting the superposition principle predicted originally by mode-coupling
theory (MCT), Fig.~\ref{alphasc}. This pattern is exhibited by all our
simulation data except for the correlators for the small particles of the
$\delta=0.60$ system for wave numbers near the structure-factor peak
position, and this for the two largest densities $\varphi=0.600$ and
$0.605$, Fig.~\ref{alphasc}. This violation of the superposition principle
 might indicate that MCT ignores relaxation processes which become
important close to the liquid-glass transition point in this mixture.
A possibility that we cannot exclude, however, are precursors of
crystallization or phase separation.

The simulation data for the alpha process have been fitted by Kohlrausch
functions and by the von~Schweidler expansion, Eqs.~\eqref{kww} and
\eqref{vonschweidler},
respectively. These fits provide two estimates of the plateau
values. Usually, these estimates agree on a $5\%$ accuracy level,
Fig.~\ref{mdkww}. Analyzing the MCT results with Kohlrausch functions,
one gets plateaus which agree with the correct values within $5\%$,
Fig.~\ref{mctkww}. This means, that both mentioned fit procedures yield
reliable estimations for the plateaus, within the indicated uncertainty
level. These plateau values exhibit a remarkable structure as functions of the
wave number $q$. There are qualitative differences in the structure of the
plateau functions referring to the $\A$ particles and the $\B$ particles.
And there are quantitative differences between the results for the two
mixtures. All these results are described quantitatively by the MCT results
except for some rare cases, where data and theory differ by up to $10\%$,
Figs.~\ref{mdampl60}--\ref{vsampl83}.

MCT requires the structure factors as input for the equations of motion.
The MCT results reported in this paper are based on the simulation results
for the studied systems, Fig.~\ref{figsq}. Replacing these structure factors
by their Percus-Yevick approximation results, MCT still reproduces all
qualitative features of the mentioned plateau functions. However, one 
systematically underestimates the data. The error of the MCT results caused
by the specified erroneous structure information can be as large as $20\%$,
Figs.~\ref{mdampl60} and \ref{mdampl83}.
These findings do not deal with
mixture-specific effects. They apply also for the simple hard-sphere system,
Fig.~\ref{fighss}.

The stretching of the alpha processes is parameterized by the Kohlrausch
exponents $\beta_\alpha(q)$, and the relative time scales are quantified by the
scales $\tau_\alpha(q)$. These quantities vary with the wave vector and they
depend on the index $\alpha$ for the species, $\alpha=\A,\B$. MCT reproduces
the $\beta_\alpha(q)$ reasonably, but there occur discrepancies up to
$20\%$. The trends for the $\tau_\alpha(q)$ are reproduced by the theory,
but there occur large quantitative errors, Figs.~\ref{figbeta} and
\ref{figtauq}.

Writing the von~Schweidler expansion, Eq.~\eqref{vonschweidler},
for the diagonal
correlators in the form $\Psi_{\alpha\alpha}(q,\tilde t)
=F_{\alpha\alpha}^c(q)\left[1-[h_\alpha^{(1)}(q)/F_{\alpha\alpha}^c(q)]
{\tilde t}^b+[h_\alpha^{(2)}(q)/F_{\alpha\alpha}^c(q)]{\tilde t}^{2b}\right]$,
one notices that $h_\alpha^{(1)}(q)$ defines a relative time scale
$\tilde\tau_\alpha(q)=[h_\alpha^{(1)}(q)/F_{\alpha\alpha}^c(q)]^{-1/b}$
while $h_\alpha^{(2)}(q)$ specifies the shape. The amplitudes
$h_\alpha^{(1)}(q)$ are reproduced reasonably by MCT, but there occur errors
up to $20\%$. MCT describes the trend of the $q$- and species-dependence
of the correction amplitudes $h_\alpha^{(2)}(q)$, but there are large
discrepancies between data and theory, Fig.~\ref{figh}.

The contradicting conclusions concerning the description of the alpha process
parameters arrived at in the preceding two paragraphs indicate that the
described problems are ones of the fitting procedures.
Indeed, Figs.~\ref{alpha-master60}
and \ref{alpha-master83} show that MCT describes the
alpha-process master functions well.

Qualitative discrepancies between MCT and simulation data concern the
ratio of the time scales $\tau(q)$ for the alpha-relaxation processes for
the density fluctuations of intermediate wave numbers $q$ and the time scale
$\tau_D$ determining the strong variation of the particle diffusivity $D$,
$D\propto1/\tau_D$. If the packing fraction of the $\delta=0.60$ system
increases from $0.51$ to $0.605$, $\log\tau(q)$ increases by $3.5$, but
$\log\tau_D$ increases only by $3.0$. The ratio $\tau(q)/\tau_D\propto
\tau(q)D$ is practically constant if $\varphi$ increases from $0.51$ to
$0.58$, i.e.\ there is perfect scale coupling within this density interval.
But increasing $\varphi$ further, the ratio increases by up to a factor $4$,
Fig.~\ref{figdtau}. MCT overestimates the trend to particle localization near
the glass transition point. This overestimation of $\tau_D$ seems to be the
reason why the calculated relaxation times $\tau(q)$ exceed the data for small
$q$, as is demonstrated in Fig.~\ref{figtauq} for $qd_\A\le5$. The longest time
scale for $q\to0$ is that for the collective diffusion process:
$\tau^{\text{coll}}(q)\propto1/(q^2D^{\text{coll}}(q))$. The coupling of
this mode to the density fluctuations causes the divergence of $\tau(q)$
for $q\to0$. One should expect that an underestimation of the tagged-particle
diffusivity $D$ implies the same mistake for the collective diffusivity
$D^{\text{coll}}$.  This explains, why the decoupling of $\tau(q)$ and
$\tau(q')$ increases if $q'$ decreases to small values, as shown by the
filled symbols in Fig.~\ref{figdtau}.

The increase of the time scales $\tau(q)$ with increasing packing fraction is
described well by the asymptotic power law for the MCT results,
Fig.~\ref{mdrect}. However, the mentioned scale decoupling implies that
the extrapolation to zero of the $(1/\tau_D)^{1/\gamma}$-versus-$\varphi$
graphs leads to an estimation of the critical packing fraction $\varphi^c$,
which exceeds the value obtained from the
$(1/\tau(q))^{1/\gamma}$-versus-$\varphi$ extrapolation by $0.6\%$
\cite{Foffi2003}.

Finally, a feature of our simulation data should be emphasized which concerns
the time regime where the correlators cross their plateaus. It deals with times
larger than the ones describing the short-time transient but preceding the
regime of validity of the alpha-relaxation scaling law. Within MCT, this
regime is described for large densities by the $\beta$-relaxation scaling
laws. In this respect, the MCT results for the hard-sphere mixtures behave
as the ones for the simple hard-sphere system \cite{Goetze2003}.
Figure~\ref{alpha-master83} shows, however, that the correlators for the
$\delta=0.83$ system are close to the plateaus for time intervals exceeding
the ones for corresponding MCT results by more than an order of magnitude.
Hence, the $\beta$-relaxation theory cannot account for the simulation
data dealing with the plateau crossing. In that respect our simulation data for
the hard-sphere mixture are also qualitatively different from the ones
measured for quasi-bidisperse hard-sphere colloids \cite{Williams2001b}
and from the simulation data for the binary Lennard-Jones mixture
\cite{Kob2003}.

\begin{acknowledgments}
W.G.\ and Th.V.\ thank their colleagues from the University of Rome for
their kind hospitality during the time this work was performed. Our
collaboration was supported in part by the European Community's
Human Potential Programme under contract HPRN-CT-2002-00307,
DYGLAGEMEM,
and the Deutsche Forschungsgemeinschaft through grant Go 154/12-2.
G.F., F.S., and P.T.\ acknowledge support from MIUR Prin and Firb and
INFM Pra-Genfdt. We thank S.~Buldyrev for providing us the simulation code
for the hard-sphere mixtures.
\end{acknowledgments}

\bibliography{mct,add}
\bibliographystyle{apsrev}

\end{document}